\documentclass[reprint, aps,amssymb,amsmath,floatfix] {revtex4-2}%

\usepackage{graphicx}
\usepackage{dcolumn}
\usepackage{bm}
\usepackage{siunitx}%
\usepackage[version=4]{mhchem}
\usepackage{xspace}
\usepackage{placeins}
\usepackage{dblfloatfix}
\usepackage{appendix}
\usepackage{lineno}
\usepackage{float}

\usepackage{makecell}
\usepackage{cleveref}

\usepackage{multirow}

\def\sNN{\ensuremath{\sqrt{s_{\mathrm{NN}}}}}
\def\pT{\ensuremath{p_{\rm T}}}
\def\pTmin{\ensuremath{p_{\rm T}^{\rm min}}}

\def\Ncoll{\ensuremath{\langle N_{\rm coll} \rangle}}
\def\Npart{\ensuremath{\langle N_{\rm part} \rangle}}
\def\vtwo{\ensuremath{v_2}}

\def\RAA{\ensuremath{R_\mathrm{AA}}}

\def\TAA{\ensuremath{\langle T_{AA} \rangle}}
\def\D0{\ensuremath{D^0}}

\def\pp{\ensuremath{pp}}
\def\AA{A\textendash A}
\def\PbPb{Pb\textendash Pb}
\def\XeXe{Xe\textendash Xe}
\def\AuAu{Au\textendash Au}

\def\CuCu{Cu\textendash Cu}

\def\pPb{\ensuremath{p}Pb}

\def\rmd{\mathrm{d}}
\def\Sloss{\ensuremath{S_{\text{loss}}}}
\def\Deltapt{\ensuremath{\Delta p_{\rm T}}}
\def\pTmin{\ensuremath{p_{\mathrm{T}}^{\text{min}}}}
\def\ebj{\ensuremath{\varepsilon_{\text{Bj}}}}
\def\ebjw{\ensuremath{\varepsilon_{\text{Bj}}^{\text{width}}}}
\def\ebji{\ensuremath{\varepsilon_{\text{Bj}}^{\text{inclusive}}}}
\def\ebje{\ensuremath{\varepsilon_{\text{Bj}}^{\text{exclusive}}}}
\def\Aperp{\ensuremath{\langle A_\perp \rangle}}
\def\Awid{\ensuremath{A_W}}
\def\Ainc{\ensuremath{A_\cup}}
\def\Aexc{\ensuremath{A_\cap}}
\def\dnchdeta{\ensuremath{\rmd N_{\text{ch}}/\rmd \eta}}
\def\eini{\ensuremath{\varepsilon_{\text{ini}}}}
\def\tauini{\ensuremath{\tau_{\text{ini}}}}
\def\sini{\ensuremath{s_{\text{ini}}}}
\def\bini{\ensuremath{b_{\text{ini}}}}
\def\Tini{\ensuremath{T_{\text{ini}}}}

\begin{document}
\title{Probing the Dependence of Partonic Energy Loss on the Initial Energy Density of the Quark Gluon Plasma}

\author{Ian Gill$^{1}$}
\affiliation{$^{1}$Wright Lab, Physics Department, Yale University, New Haven, CT 06520, USA}
\author{Ryan J. Hamilton$^{1}$}
\email[Electronic address: ]{ryan.hamilton@yale.edu}
\affiliation{$^{1}$Wright Lab, Physics Department, Yale University, New Haven, CT 06520, USA}
\author{Helen Caines$^{1}$}
\affiliation{$^{1}$Wright Lab, Physics Department, Yale University, New Haven, CT 06520, USA}

\date{\today} 

\begin{abstract} 
\noindent Considerable evidence now exists for partonic energy loss due to interaction with the hot, dense medium created in ultra-relativistic heavy-ion collisions. A primary signal of this energy loss is the suppression of high transverse momentum \pT\ hadron yields in \AA\ collisions relative to appropriately scaled \pp\ collisions at the same energy. Measuring the collision energy dependence of this energy loss is vital to understanding the medium, but it is difficult to disentangle the medium-driven energy loss from the natural kinematic variance of the steeply-falling \pT\ spectra across different collision center of mass energy per nucleon pair \sNN. To decouple these effects, we utilize a phenomenologically motivated spectrum shift model to estimate the average transverse momentum loss \Deltapt\ imparted on high \pT\ partons in \AA\ collisions, a proxy for the medium induced energy loss. 
We observe a striking correlation between \Deltapt\ and Glauber-derived estimates of initial state energy density \ebj, consistent across two orders of magnitude in collision energy for a variety of nuclear species. To access the path-length dependence of energy loss, we couple our model to geometric event shape estimates extracted from Glauber calculations to produce predictions for high-\pT\ hadron elliptic flow \vtwo\ that agree reasonably with data.
\end{abstract}
\maketitle

\section{Introduction}
\label{sec:Introduction}
Over the past two decades, strong evidence has accumulated for the creation of the Quark Gluon Plasma (QGP) in collisions of relativistic heavy-ions through extensive study at both the Relativistic Heavy-Ion Collider (RHIC) at Brookhaven National Laboratory, NY, USA and the Large Hadron Collider (LHC) at CERN in Geneva, Switzerland (see, for example, Ref.~\cite{Harris:2024aov} and the references therein). Forefront among this evidence is partonic energy loss, termed ``jet quenching." This is revealed via the suppression of high transverse momentum (high-\pT) hadrons in nucleus-nucleus (\AA) collisions relative to their production in \pp\ collisions scaled by the mean number of binary nucleon-nucleon collisions \Ncoll\ so as to make the two comparable. Experimentally, this suppression is frequently identified via the nuclear modification factor
\begin{equation}
    R_{AA} = \frac{1}{\TAA}\frac{\rmd^{3}N_{\rm ch}^{AA}/\rmd p_{T} \rmd\eta \rmd \phi}{\rmd^3\sigma_{\rm ch}^{pp}/\rmd p_{T} \rmd\eta \rmd \phi},
\end{equation}
where $\TAA = \Ncoll/\sigma_{\text{inel}}^{\text{NN}} $ is the nuclear overlap function determined from Glauber model calculations~\cite{Loizides:2017ack}, proportional to \Ncoll\ via the inelastic nucleon-nucleon cross section $\sigma_{\text{inel}}^{\text{NN}}$ at the relevant center of mass energy per colliding nucleon pair $\sqrt{s_{\text{NN}}}$. $N_{\text{ch}}^{AA}$ and $\sigma_{\text{ch}}^{\pp}$ denote the charged particle yield per event in \AA\ collisions and the charged particle production cross section in \pp\ collisions, respectively. 

The observation of \RAA\ below unity indicates that partons (quarks and gluons) lose energy as they traverse the dense medium created in the collision~\cite{Harris:2024aov}; data consistently show this signal across the \pT\ range but especially at high\textendash \pT, where the effect is attributed to jet quenching. Centrally, understanding the forces driving this partonic energy loss necessitates understanding the collision energy dependence. While the measured \RAA\ suppression values for different collision \sNN\ are comparable in principle, \RAA\ results contain a variety of convolved effects: medium-driven effects like collectivity and jet quenching, as well as kinematic restrictions of initial state parton composition and spectral shape. \RAA\ is strictly a ratio of yields, but is identified with energy loss because a uniform change in \pT\ does not uniformly affect \pT-differential yields for steeply falling \pT\ spectra. LHC \pT\ spectra, reflecting higher collision energy, are less steeply falling and dominated by gluon fragmentation when compared to the significantly softer RHIC spectra which primarily originate from the fragmentation of quarks \cite{Eskola:2004cr, Harris:2024aov}. Hence, similar nuclear modification factors \RAA\ at RHIC and the LHC actually indicate more energy loss for partons traversing QGP created at the LHC. Additionally, observed differences between the \RAA\ for photon-tagged jets and inclusive jets, such as that reported by ATLAS in \sNN\ = 5.02 TeV \PbPb\ collisions~\cite{ATLAS:2023iad}, indicate that quarks and gluons interact differently with the QGP. These results highlight the importance of reliably deconvoluting medium effects from kinematic ones; the inherent hadron spectrum slope difference between RHIC and LHC alone can propagate to as much as a 10\% difference in the \RAA. 

The abundance of available jet quenching data from experimental collaborations such as high-\pT\ \RAA\ measured below unity, alongside the challenge of calculations in QCD and related Effective Field Theories (EFTs), has motivated a wealth of phenomenological work attempting to bridge experimental observation with theoretical results. A non-exhaustive list of studies include EFT-motivated frameworks of jet quenching~\cite{Djordjevic:2008iz, Djordjevic:2009cr, Zapp:2012ak, JET:2013cls, Spousta:2015fca, Andres:2016iys, Arleo:2017ntr, Zigic:2018ovr, Zigic:2018smz, Mehtar-Tani:2021fud, Zigic:2021rku, Karmakar:2023ity, Karmakar:2024jak, Faraday:2024gzx, Faraday:2024qtl, Pablos:2025cli} and corresponding path-length dependent energy loss~\cite{Zigic:2018ovr, Zigic:2018smz, Andres:2019eus, Wu:2023azi, Arleo:2022shs, Ogrodnik:2024qug, Mehtar-Tani:2024jtd, Karmakar:2024jak}, Bayesian extractions~\cite{He:2018gks, Zhang:2022rby, Xing:2023ciw, Zhang:2023oid,  Wu:2023azi, Ehlers:2024miy, Falcao:2024zkw, Pablos:2025cli}, Boltzmann transport models~\cite{He:2018gks, He:2018xjv, Zhang:2021oki, Zhang:2022rby, Zhang:2023oid, Xing:2023ciw, Ogrodnik:2024qug, Soltz:2024gkm}, quenched jet substructure~\cite{Chien:2018dfn, Ehlers:2024miy, Chen:2019gqo}, gauge boson-tagged phenomenology~\cite{Neufeld:2010fj, Zhang:2021oki}, and modern computational techniques~\cite{Brewer:2020och,Chien:2018dfn, He:2018gks, Takacs:2021bpv, Pablos:2025cli}; often a combination of these approaches is employed. For reviews on jet quenching and methods more broadly see~\cite{Mehtar-Tani:2013pia, Qin:2015srf, Cao:2024pxc, Paquet:2023rfd}, for reviews on phenomenology in jet quenching see~\cite{Cao:2024pxc, Majumder:2010qh}.

On the experimental side, the effort to disentangle medium signatures from kinematic effects and better interpret \RAA\ results has led several collaborations to study the shift in \pT\textemdash first proposed as \Sloss\ by PHENIX~\cite{PHENIX:2004vcz}\textemdash needed to align the spectra in \AA\ with the binary scaled \pp. Subsequent theoretical and phenomenological studies of jet quenching have also explored similar measures of energy loss related to horizontal shifts of \pT\ spectra~\cite{Brewer:2018dfs, Spousta:2015fca, Arleo:2017ntr, He:2018xjv, He:2018gks, Falcao:2024zkw, Soltz:2024gkm, Mehtar-Tani:2021fud, Takacs:2021bpv}; we will refer to this strategy broadly as \Deltapt\ methods. Unlike \RAA, these \Deltapt\ measures perform a direct fitting between hadron spectra, significantly reducing effects of \pT\ distribution shape on the results. Measurements of \Sloss\ and other similar \Deltapt\ observables show that the average energy loss at LHC energies exceeds that observed at RHIC, and by extension that gluon-dominated systems exhibit more energy loss than quark-dominated ones~\cite{PHENIX:2015vqa, ATLAS:2023iad, Sahoo:2020kwh, PHENIX:2012jha, PHENIX:2012oed, PHENIX:2006wwy, PHENIX:2004vcz}.

The goal of this analysis is to determine the degree of correlation between the medium-induced partonic energy loss and the initial energy density of the QGP. To quantify energy loss, we extract the average \Deltapt\ from experimentally observed charged particle spectra at high \pT\ in \pp\ and \AA\ collisions. This work differs from previous studies by assuming a fixed \Deltapt\ with the aim of estimating the average energy loss for a given centrality range and collision energy. While more rigorous first-principle calculations indicate that a fractional partonic energy loss is potentially preferred~\cite{Baier:2001yt}, the limited \pT\ ranges studied in this work make our fixed \Deltapt\ approximation reasonable. In addition, given that high \pT\ tracks, instead of full jets, are used as in this study to approximate the initial parton energy loss it is not clear that a fractional energy loss signature will be maintained through the natural smearing of the correlation between partonic energy and hadron \pT\ that occurs during the fragmentation and hadronization process. 

For QGP energy density, we use the reported particle yields at mid-rapidity and Glauber calculations to estimate the initial state energy density in the limit of Bjorken flow \ebj. By further utilizing Glauber event geometry, we predict the hadronic high-\pT\ \vtwo\ assuming a linear path-length dependent energy loss. The \vtwo\ serves as an additional arena to compare degrees of freedom in our model and probes the relation between medium path length and energy loss.

This article is organized as follows: Section~\ref{sec:Analysis} details the data analysis including the determination of the transverse overlap area of the collisions (\ref{subsec:Aperp}), the Bjorken energy density estimation (\ref{subsec:ebj}), the techniques used to extract \Deltapt\ (\ref{subsec:Deltapt}) and the high-\pT\ \vtwo\ predictions (\ref{subsec:v2}). Section~\ref{sec:Results} presents our results. We conclude with a discussion in Section~\ref{sec:Conc} that summarizes our findings.

\section{Analysis}\label{sec:Analysis}

The analysis proceeds along three axes. First, estimates of the initial transverse overlap area of the two colliding nuclei, \Aperp, are made using Monte Carlo Glauber calculations~\cite{Loizides:2017ack}. The initial energy density is then approximated using \Aperp\ and the charged particle multiplicities \dnchdeta\ that have been determined experimentally. Second, the \Deltapt\ of each dataset is determined from the reported charged particle \AA\ and \pp\ transverse momentum spectra. Once both the energy density and \Deltapt\ values have been calculated for each centrality bin of each collision system, the correlation between these quantities is determined. Finally, the initial overlap geometry is estimated from the Glauber simulation. Geometric quantities together with \Deltapt\ enable prediction of the high-\pT\ \vtwo\ to be made for each dataset, assuming a linear path-length dependent energy loss.

The principle data sets used in this analysis are \XeXe\ charged particle spectra from ALICE~\cite{ALICE:2018hza} and ATLAS~\cite{ATLAS:2022kqu} at \sNN\ = 5.44 TeV, ALICE~\cite{ALICE:2018vuu} \PbPb\ results at \sNN\ = 5.02 TeV and 2.76 TeV, STAR~\cite{STAR:2003fka} and PHENIX~\cite{PHENIX:2003djd} charged particle spectra from \AuAu\ collisions at \sNN\ = 200 GeV, and STAR $\pi^{\pm}$ data from \CuCu~\cite{STAR:2009mwg} also at \sNN\ = 200 GeV. Corresponding \pp\ spectra were obtained from the same \AA\ references for all datasets except STAR \AuAu, for which 200 GeV \pp\ spectra were found in Ref.~\cite{STAR:2011iap}. All data were obtained from publicly available HEPData archives. 

\begin{figure}[!ht]
    \centering
    \includegraphics[width=0.50\textwidth]{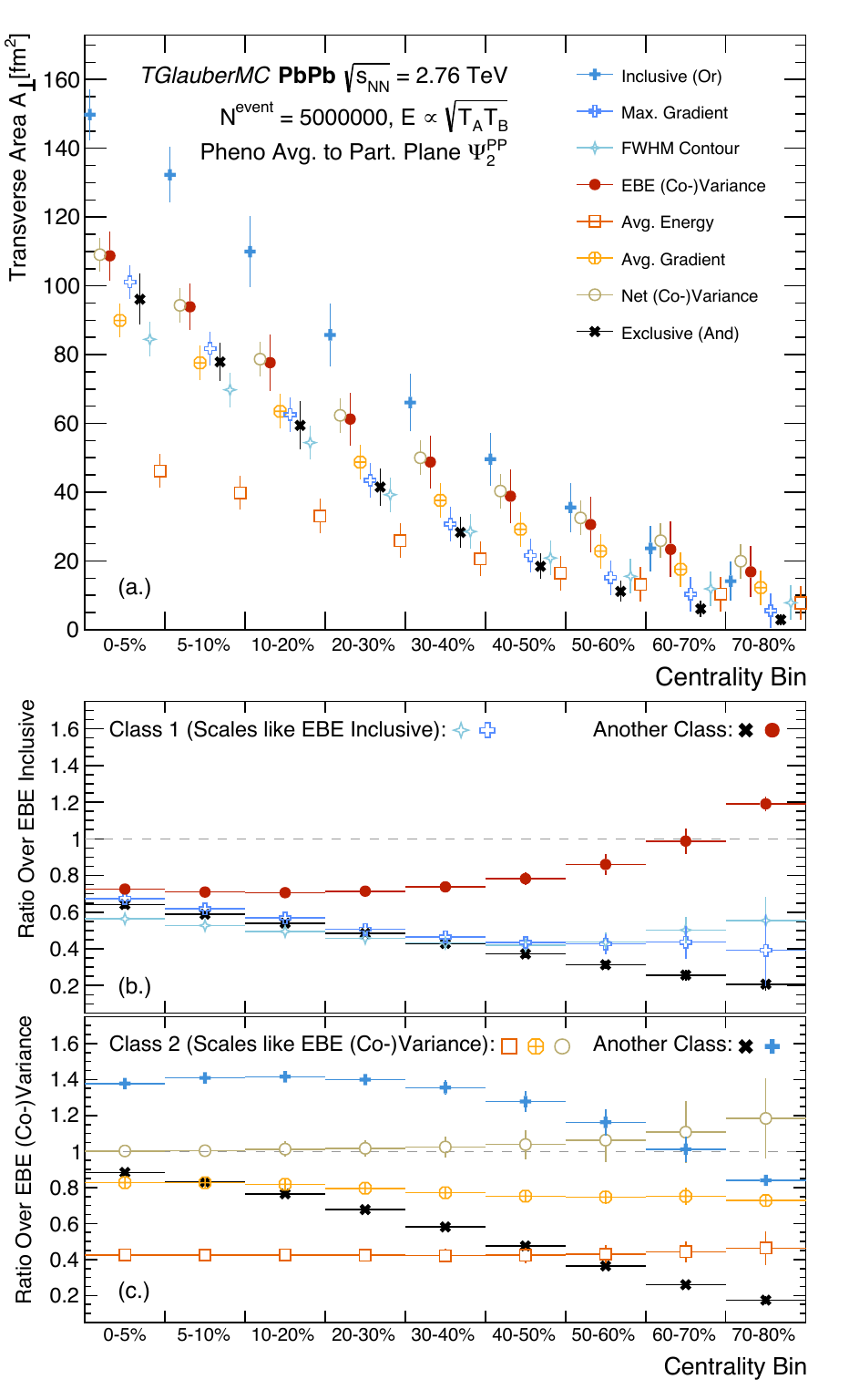} 
    \caption{(a.) Computed transverse areas as a function of the event centrality for \PbPb\ collisions at \sNN\ = 2.76 TeV. Ratios of certain methods against (b.) \Ainc\ and (c.) \Awid. The blue and orange markers denote the class of methods that scale like \Ainc\ and \Awid\ respectively. The EBE exclusive calculations are shown as the black cross markers. See text for details.} 
    \label{fig:areascale_PbPb2.76} 
    \label{subfig:areascale_PbPb2.76_a} 
    \label{subfig:areascale_PbPb2.76_b}
    \label{subfig:areascale_PbPb2.76_c}
\end{figure}

\subsection{Transverse Area Estimation} \label{subsec:Aperp}

The transverse overlap area of the collision \Aperp\ must be determined from simulation, and there is currently no definitive technique to perform this extraction. We therefore start this study by determining \Aperp\ using several physically reasonable approaches, with the goal of determining if any exhibit similar behavior. First, we utilize the three grid-based \Aperp\ calculations directly available from the Glauber code and described in Ref.~\cite{Loizides:2016djv}. We then present four new \Aperp\ determinations based on averaged initial collision energy deposition. 
 
The three built-in Glauber model~\cite{Loizides:2016djv,Loizides:2017ack} estimates of transverse area\textemdash width-based, inclusive, and exclusive areas\textemdash are calculated on an event-by-event basis. The statistical width-based area, \Awid, is calculated from the (co-)variances of the nucleon distributions; the inclusive area, \Ainc, is a grid-based set union including a disk around all participant nucleons; the exclusive area, \Aexc, is a grid-based set intersection including only regions struck by both nuclei. The uncertainty on each area calculation is determined by the standard error of the mean of the output of the Glauber calculations. We label these three area calculations the event-by-event (EBE) methods.

The EBE methods are compared against estimates extracted from an averaged initial state energy density distribution $E(x,y)$ in the transverse plane. In each event, the Glauber-generated nuclear thickness profiles are used to calculate local energy density via the geometric-mean scaling $E \propto \sqrt{T_A T_B}$. This energy scaling enables the Glauber Monte Carlo to produce initial state energy distributions that agree qualitatively with more modern models like IP-Glasma \cite{Schenke:2012wb}, and is perhaps the Glauber energy scaling best supported by data~\cite{Bernhard:2019bmu}. The resulting single-event transverse energy profiles were then translated and rotated to align the center-of-mass position and the second-order participant plane angle $\Psi_2$. Finally the profiles were averaged to produce a single representative initial-state energy distribution in each centrality bin. Following established Glauber procedures~\cite{Loizides:2017ack}, centrality was determined at simulation level by binning in the impact parameter $b$ as a proxy for final-state multiplicity. To reconcile this procedure with Glauber simulations used by the collaborations, which determine centrality by coupling Glauber Monte Carlo calculations to a simulation of a physical detector response~\cite{PHENIX:2003djd,ATLAS:2022kqu, ALICE:2018vuu, ALICE:2018hza, ALICE:2018tvk} or otherwise bin in \Ncoll\ itself~\cite{STAR:2003fka, STAR:2002ggv, STAR:2009mwg,PHOBOS:2002lqa}, a systematic error is included on \Ncoll\ as follows. The Glauber Monte Carlo~\cite{Loizides:2017ack} is run with a variation of inelastic cross section with uncertainty $\sigma_{\rm inel}^{\rm NN} \pm \delta \sigma_{\rm inel}^{\rm NN}$ as reported by the collaboration, and also with nuclear shape parameters $\beta_{2, 4}, \gamma$ disabled. The \Ncoll\ values reported by the collaboration are also tabulated. To obtain a systematic uncertainty on \Ncoll, we compare the nominal Glauber \Ncoll\ results\textemdash for which $\sigma_{\rm inel}^{\rm NN}$ is always obtained from Ref.~\cite{Loizides:2017ack}\textemdash against each of the cases mentioned above. The largest difference $\delta\Ncoll$ against any variation in each centrality bin is symmetrized and taken as the systematic error on \Ncoll\ for that respective centrality bin. This error is used in the full procedure and extrapolated to a final uncertainty on the results. 

Using this representative initial-state energy distribution, the average transverse area in a centrality bin was computed by extracting an ``edge" azimuthal function $R(\phi)$. The area contained within this curve is the transverse area. Four methods were considered to determine $R(\phi)$: the average energy radius 
\begin{equation}\label{eq:R_average_radius}
 R_{\langle E \rangle}(\phi) = \frac{1}{E_0}\int_0^\infty r E(r, \phi) \, \mathrm{d}r,
\end{equation}
the average pressure (energy gradient) radius 
\begin{equation}\label{eq:R_average_gradient}
    R_{\langle P\rangle}(\phi) = \frac{1}{\delta E_0}\int_0^\infty r \left|\nabla E(r, \phi)\right| \, \mathrm{d}r.
\end{equation}
For simplicity, we choose $E_0 = 1$ GeV/fm, $\delta E_0 = 1$ GeV / fm$^2$; this can be thought of as translating the energy density maps to probability density distributions as in the Optical Glauber formalism. In addition, we consider the full-width-at-half-max contour which solves
\begin{equation}\label{eq:R_half_max_countour}
    E(R_{\text{FWHM}}(\phi), \phi) = \frac{E_{\text{max}}}{2},
\end{equation}
and the surface of maximal gradient/pressure 
\begin{equation}\label{eq:R_maximal_gradient}
    R_{P, \text{max}}(\phi) = \underset{r \ge 0}{\operatorname{arg\,max}}\left|\nabla E(r, \phi)\right|.
\end{equation}
Lastly, we also computed the (co-)variance ``width'' based area of the representative distribution, as was done on the event-by-event level. Confirming that the EBE \Awid\ and net-event averaged (co-)variance width areas agree is a validation of the event averaging procedure. These five methods are collectively labeled phenomenological areas, since while each edge-extraction method can be physically motivated (e.g. the average radius method $R_{\langle E\rangle}(\phi)$ represents the average total energy seen by a traversing high energy parton emitted at angle $\phi$), it is not immediately clear which method, if any, might best represent the effective \Aperp\ relevant to this study. 

The abundance of methods available for computing the transverse area motivates us to categorize them and search for defining properties of a given method's predictions. For the purposes of this analysis, the primary behavior of interest is the centrality dependence of the transverse area, modulo any overall normalization in that dependence. For an illustrative example, we consider the centrality dependence of the various Glauber areas for \PbPb\ at \sNN\ = 2.76 TeV, shown in Fig. \ref{subfig:areascale_PbPb2.76_a}(a), for each of the methods described above. 

A trend emerges when considering certain ratios. The eight transverse area calculations can be grouped into two classes and one outlier \Aexc. The methods are considered as belonging to the same ``class'' if the ratio of their centrality dependences is roughly flat. Throughout Fig.~\ref{fig:areascale_PbPb2.76}, consistent hue and marker shapes are used to denote class membership among methods. We denote the two classes as the inclusive class \Ainc\ and the width \Awid\ class for methods that scale like the EBE inclusive area and EBE width-based calculations respectively. Figure~
\ref{subfig:areascale_PbPb2.76_b}(b) shows ratios against the EBE inclusive \Ainc\ calculation, where the phenomenological Maximum Gradient and Full-Width at Half-Max contour methods are approximately flat. These two methods form the inclusive class together with \Ainc. Figure~ \ref{subfig:areascale_PbPb2.76_c}(c) shows ratios against the EBE width \Awid\ method, where the remaining three phenomenological methods\textemdash the average energy radius, average gradient radius, and (co-)variance width\textemdash exhibit flat ratios, forming the width class \Awid. Note that the pre-averaged EBE width calculation \Awid\ and the post-averaged phenomenological width calculation exhibit ratios consistent with unity, a check on the event averaging procedure. In each case, \Aexc\ exhibits markedly different scaling, shown by non-constant ratios. The EBE inclusive \Ainc\ and width \Awid\ also do not have consistent scalings, meaning they cannot be merged to a single class. 

While initially disconcerting, the disagreement between \Aexc\ and other methods also coincides with observations in ALICE \pPb, where computations using \Aexc\ produce unusually high estimates of energy density~\cite{ALICE:2022imr}. Physically, the \Aexc\ and \Ainc\ methods reflect distinct pictures of Glauber modeling: \Ainc\ considers entire struck nucleons as a participants in traditional Glauber fashion, while \Aexc\ regards only overlapping sub-nucleonic regions of participant nucleons as contributing. 

Lastly, it is important to note that while Fig. \ref{fig:areascale_PbPb2.76} shows the calculations for \PbPb\ collisions at \sNN\ = 2.76 TeV, the division of methods into these three distinct classes is persistent across all the energies and species studied, as detailed in Appendix \ref{appendix:data_supplement}.

\subsection{Bjorken Energy Density Estimation} \label{subsec:ebj}

The average initial energy density of the medium \eini\ as a function of centrality is approximated in the limit of Bjorken hydrodynamics~\cite{Bjorken:1982qr} as outlined in Ref.~\cite{Harris:2024aov}; this results in the approximation
\begin{equation}
    \eini \sim \ebj \approx \frac{3}{4} \left( \frac{7J\frac{\mathrm{d}N_{\text{ch}}}{\mathrm{d}\eta}}{\Aperp b_{\text{ini}} \tau_{\text{ini}}} \right)^{\frac{1}{3}} \frac{7J \frac{\mathrm{d}N_{\text{ch}}}{\mathrm{d}\eta}}{\Aperp \tau_{\text{ini}}},
\end{equation}
where $J$ is the Jacobian conversion from pseudorapidity to rapidity, \dnchdeta\ is the charged hadron density at midrapidity, \tauini\ is the QGP formation time, and \bini\ is defined by $\sini = \bini\Tini^3$; \sini\ and \Tini\ are the initial entropy density and temperature respectively. As in Ref.~\cite{Harris:2024aov}, \bini\ is assumed to be constant and independent of collision energy, species, and centrality with a value of \bini\ = 15.5. For highly Lorentz contracted nuclei the parameter \tauini\ is commonly chosen to be 0.6 fm/$c$, while at collision energies $\sNN \lesssim 20$ GeV, the time for the nucleons to fully cross is longer~\cite{Harris:2024aov}. As all collision data used in this study have center-of-mass energy per nucleon $\sNN \geq 200$ GeV, we also fix \tauini\ = 0.6 fm/$c$.

The remaining input parameters were obtained from various sources. The Jacobian $J$ ranges from 1.25 for \sNN\ = 200 GeV to 1.09 at \sNN\ = 5.44 TeV~\cite{PHENIX:2015tbb,CMS:2012}. An uncertainty of 3\% on $J$ is assumed for all beam energies, propagated forward to the uncertainty on \ebj. The \dnchdeta\ data used in this analysis are those reported from \PbPb\ collisions at \sNN\ = 2.76 TeV and \sNN\ = 5.02 TeV from ALICE~\cite{ALICE:2010mlf, ALICE:2018vuu}, \XeXe\ collisions at \sNN\ = 5.44 TeV from ALICE~\cite{ALICE:2018hza}, \AuAu\ collisions at \sNN\ = 200 GeV from STAR~\cite{BRAHMS:2001llo}, and lastly \CuCu\ collisions at \sNN\ = 200 GeV from STAR~\cite{STAR:2009mwg}. The uncertainty used on \dnchdeta\ is the quadrature sum of the reported statistical and systematic uncertainties in each measurement. 

The transverse area \Aperp\ is determined for each species, beam energy, and corresponding centrality class as described in the previous section. Uncertainties originating from \dnchdeta\, \Aperp\, and $J$ are propagated differentially to the final energy density value. To fairly consider each of the three observed centrality scalings, we separately compute energy densities from each of the EBE methods with \ebjw, \ebji, \ebje\ representing width-based \Awid, inclusive \Ainc, and exclusive \Aexc\ areas respectively. Details of the resultant values and uncertainties of \ebjw\ and \ebji\ computed from the above inputs are given in Appendix~\ref{appendix:data_supplement}, along with the relevant data used to compute these quantities. The exclusive area \ebje\ is not included in the tables, as its especially sharp centrality scaling seen in Fig.~\ref{fig:areascale_PbPb2.76} leads to unphysical trends where the energy density \ebje\ is smaller in central collisions than peripheral ones for some systems; unexpected behaviors of the exclusive method have also been observed by ALICE in LHC \pPb\ collisions~\cite{ALICE:2022imr}, where \Aexc\ predicts higher energy density in \pPb\ collisions than in central \PbPb. Still, to enable a fair comparison among all methods, we do include \ebje\ in the final results of Sec.~\ref{sec:Results}.

\subsection{Extracting \Deltapt\ from Transverse Momentum Spectrum Data} \label{subsec:Deltapt}

In previous analyses~\cite{PHENIX:2015vqa,ATLAS:2023iad, Sahoo:2020kwh,Brewer:2018dfs,PHENIX:2012jha,PHENIX:2012oed,PHENIX:2006wwy,PHENIX:2004vcz} \Deltapt\ was determined for a fixed hadron \pT. Using the transverse momentum spectra found in~\cite{ALICE:2018vuu,ALICE:2018hza,PHENIX:2003djd,STAR:2003fka,STAR:2011iap,STAR:2009mwg,ATLAS:2022kqu}, this study explores if a common \Deltapt\ can be identified for the whole range of high-\pT\ data reported. We identify such a \Deltapt\ via a \pT\ spectrum shifting procedure described in this section.

The horizontal shifting of the \pp\ \pT\ spectra to describe mean parton energy loss is applicable only at high momentum, where parton fragmentation is the dominant source of particle production. A threshold \pT\ value, \pTmin, must therefore be chosen to ensure that other collective effects\textemdash such as radial flow\textemdash present at lower \pT\ do not affect the extraction. Data below this \pT\ threshold are not included in the determination of \Deltapt. To determine this threshold we consider the reported charged hadron \RAA. In all measurements studied, the \RAA\ in central collisions exhibit a local minimum near \pT\ $\sim 5$ GeV/c, whereafter the \RAA\ data rise monotonically. While jet quenching effects are likely to still be present at momenta below this turnover, the region above should be reasonably free of collective effects; we therefore take these local minima as \pTmin. The systematic uncertainty on \Deltapt\ due to this choice of threshold is determined by varying the \pT\ threshold by an additional \pT\ bin below and above each dataset's local \RAA\ minima.
\begin{figure}
    \centering
    \includegraphics[width=\linewidth]{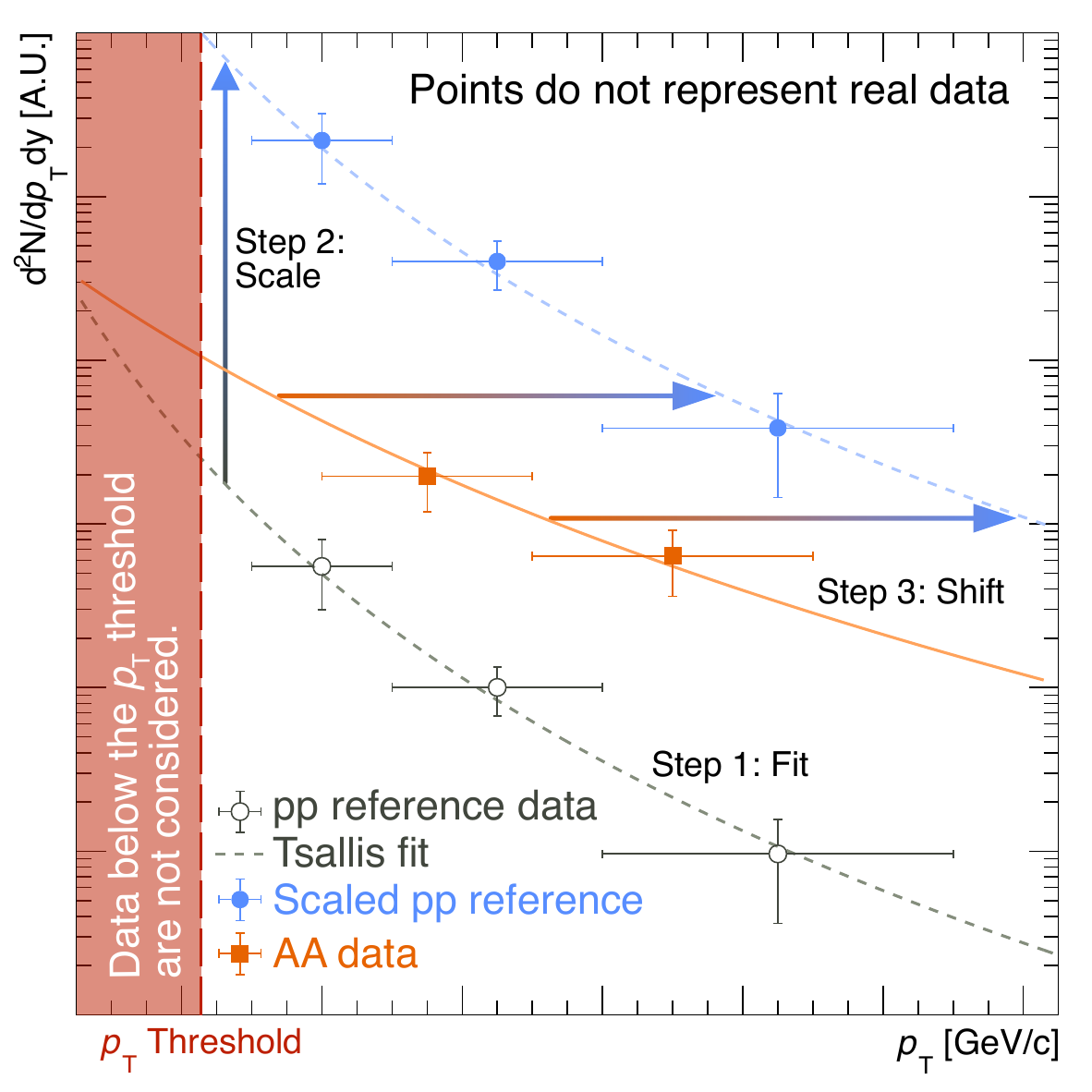}
    \caption{Cartoon illustrating the procedure used to determine \Deltapt.}
    \label{fig:process_pT}
\end{figure}

Since many of the peripheral datasets suffer from large uncertainties, preventing a clear determination of a local minimum, we decided to use the \pTmin\ extracted in the most central events for all centralities reported for a given dataset. This likely results in a slight overestimate of the \pT\ threshold in peripheral events since glancing collisions should produce a narrower collective region. As we are aiming to exclude collective phenomena, this overestimation should not affect our conclusions. 

\par With the \pT\ threshold chosen, we compute the \Deltapt\ by scaling and shifting the \pp\ reference spectra to match the \AA\ data using the following procedure:
\begin{enumerate}
    \item Reference \pp\ spectra are first fit to a Tsallis distribution, given by
    \begin{equation}
        \frac{1}{2\pi} \frac{\rmd^2 N}{\rmd p_{\rm T} \rmd \eta} \sim E\frac{\mathrm{d}^3 N}{\mathrm{d}p^3} = C \left(1 + \frac{E_{\text{T}}}{nT}\right)^{-n},
        \label{eq:Tsallis} 
    \end{equation}
    where $E_{\rm T} = \sqrt{m^2 + p_{\rm T}^2} - m$ is the transverse kinetic energy, $n$ is the high-\pT\ power law scaling of the \pT\ spectrum, $T$ is a temperature-like parameter controlling the width of the collective region, and $C$ is a normalization factor. We use the standard pion mass $m=m_\pi$ in the kinetic energy $E_{\rm T}$ for all particles. Motivation for this choice of fit function and details about our fitting procedure will be discussed shortly. 
    \item The resultant fit on the \pp\ spectra is then scaled by a factor \TAA\ obtained from MC Glauber estimates \cite{Loizides:2017ack}, as they would be for an $\RAA$ calculation.
    \item The \AA\ data are shifted horizontally rightward toward high-\pT\ (or equivalently, the scaled \pp\ fit can be translated horizontally leftward) until the shifted \pp\ baseline spectrum and \AA\ data agree as well as possible, according to the same fit metric used for the \pp\ spectrum fitting. This optimal horizontal shift is \Deltapt.
\end{enumerate}
This procedure is shown diagrammatically in Fig.~\ref{fig:process_pT}.

The fit methods for the \pp\ spectra and \Deltapt\ shift were chosen carefully to best extract the relevant parameters for our analysis: namely the power law scaling of the various \pT\ spectra. The Tsallis distribution, also called the Hagedorn function, was chosen as the fit form following other work demonstrating that this function is an effective choice for \pp\ spectra over the full \pT\ range at both RHIC and LHC energies~\cite{Wong:2012zr,Wilk:1999dr,ATLAS:2010jvh,ALICE:2013txf,STAR:2006nmo,PHENIX:2011rvu}. The function smoothly connects a thermodynamic, low-\pT\ region expected to scale as $\sim e^{-\pT}$ with the observed power law scaling at high \pT. Since data below \pTmin\ are excluded from the fit, in principle any function with power law asymptotics should yield comparable \Deltapt, but a known effective form was chosen as a safeguard. Fitting of the \pp\ spectra is performed by binning a candidate Tsallis fit and computing a fit metric against the binned data; the set of parameters which minimize the metric are selected as the final fit. We found that the commonly used $\chi^2$ fit metric was reasonable but tended to be heavily biased by the first bin above the \pT\ threshold, which can be many orders of magnitude larger than bins at higher \pT, and therefore generally produced poor extractions of the power law, even in toy fits with infinite statistics. To avoid this bias, we selected a fit metric which compares logarithms of the bin contents. We found such a metric extracted the power law with significantly higher accuracy in all tests we performed; for details see Appendix~\ref{appendix:fitting_methods}. The metric we chose takes the form
\begin{equation}\label{eq:MSE_definition}
    \text{MSE} \equiv \sum_i \left( \log\frac{O_i}{E_i}\right)^2,
\end{equation}
where $O_i$ and $E_i$ are the data and fit candidate; we label this metric the Mean Square Entropy (MSE). The form has some similarity to G-tests or likelihood tests in statistics. The use of this metric is analogous to performing a linear regression in the log-log plane, which allows it to reliably extract precise estimates of the power law scaling. For consistency, we used this fit metric for both the \pT\ spectra fitting and the \Deltapt\ shift fitting. 

In addition to the systematic uncertainty generated from the choice of $p_T^{\text{min}}$, uncertainties in both \pp\ and \AA\ \pT\ spectra are propagated to the uncertainty on \Deltapt\ using a Monte Carlo method. Individual bins of the \pT\ spectra are randomly varied using normal distributions with widths that matched the uncertainties, and \Deltapt\ is recalculated using these smeared spectra. The mean of 10000 variations is reported as the final \Deltapt\ shift, and the standard deviation becomes the propagated uncertainty. Uncertainty on \Ncoll\ described in Sec.~\ref{subsec:Aperp} is propagated to the \Deltapt\ by performing the \pp\ scaling component of the shifting procedure with \Ncoll\ replaced by  \Ncoll$-\delta$\Ncoll\ and \Ncoll$+ \delta$\Ncoll. The maximum difference between the default  \Deltapt\ and these altered values is taken as the uncertainty from this source. All uncertainty sources are added in quadrature to obtain the final \Deltapt\ uncertainty. 

\subsection{High \pT\ \vtwo\ Estimation} \label{subsec:v2}
\begin{figure}[b]
    \centering
\includegraphics[width=\linewidth]{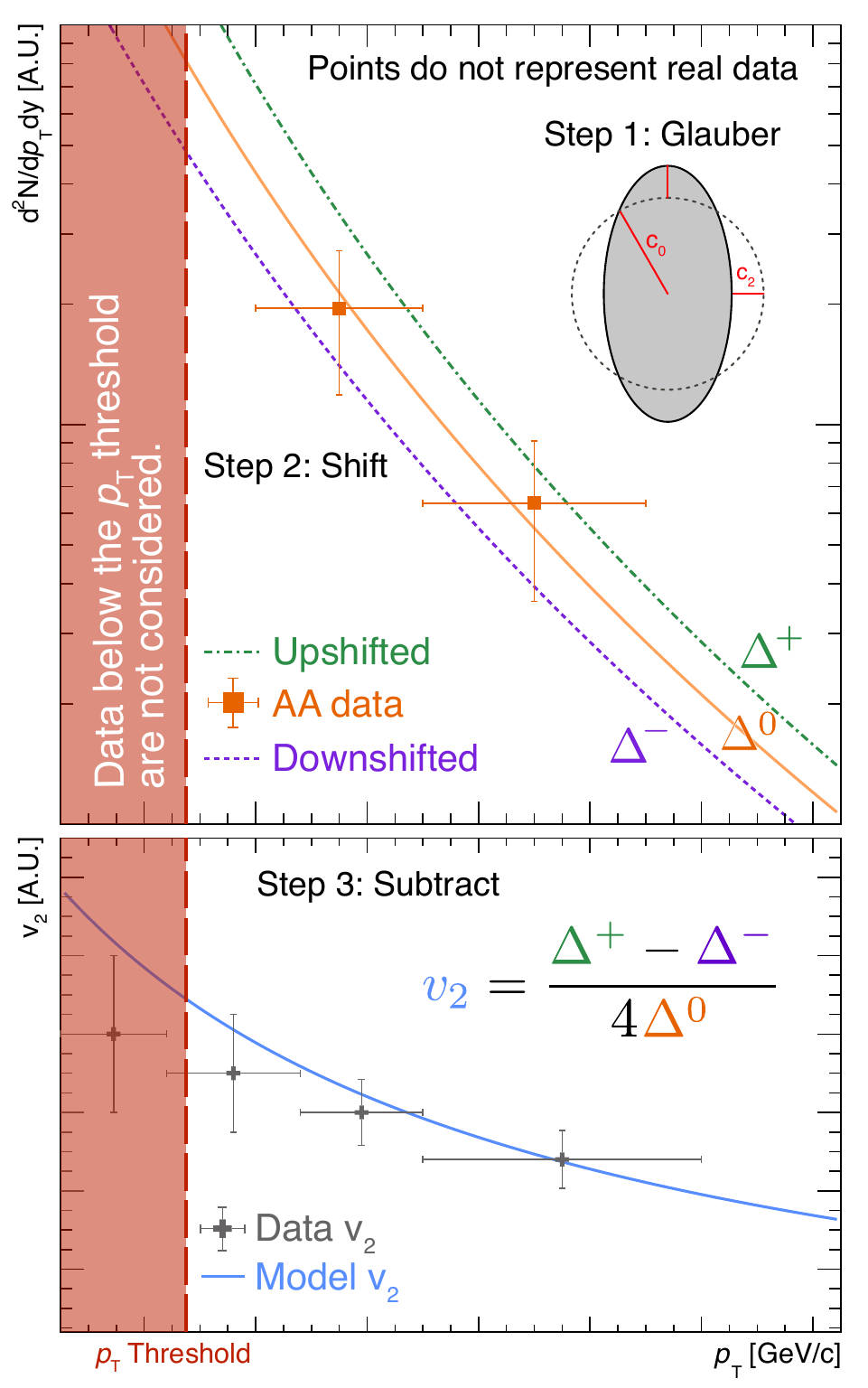}
    \caption{Cartoon illustrating the steps of the \vtwo\ estimation process.}
    \label{fig:process_v2}
\end{figure}
As described above, the first part of this study approximates the initial energy density from the mid-rapidity hadron multiplicity \dnchdeta\ and Glauber estimates of the initial geometry and overlap area \Aperp. However, the correlation between these quantities is complex and merits further exploration; the link between energy density and charged particle multiplicity is expected to be sensitive to the medium shear viscosity $\eta$~\cite{Giacalone:2019ldn}, and the relationship between energy loss and event geometry informs the path-length dependence of medium-driven energy loss. We therefore turn to \pT-differential azimuthal anisotropy. 

\par The observed \pT\ dependent hadron multiplicity as a function of azimuthal angle admits a Fourier series decomposition, 
\begin{equation}\label{eq:azimuthal_anisotropy_fourier_decomp}
    \frac{\rmd^2N}{\rmd p_{\mathrm{T}} \rmd\Delta_i\phi} = \frac{1}{2\pi} \frac{\rmd N}{\rmd \pT} \left[1 + 2\sum_{n=1}^{\infty} v_n(p_{\mathrm{T}}) \cos(n \Delta_i \phi)\right],
\end{equation}
where the angle $\Delta_i\phi = \phi - \Psi_i$ is oriented relative to the $i^{\rm th}$-order collision event plane, measured on an event-by-event basis. $\phi$ is the particle's azimuthal angle, and $\Psi_i$ is that of the plane. Roughly speaking, the magnitude of the extracted high-\pT\ \vtwo\ has two distinct contributions at each \pT: one from higher energy initial partons that lost more energy traversing a longer length of QGP, and another from lower energy initial partons that lost less energy traversing a shorter length. After determining the average \Deltapt\ for a given centrality as described above, we can use this principle to construct a model for estimating the charged particle high \pT\ $\vtwo(\pT)$ as follows:
\begin{enumerate}
    \item Consider a Glauber derived estimate of the path length of an event-averaged collision region as a function of azimuth $R(\Delta\phi)$, discussed above as phenomenological area estimates. Fourier decompose this function, and take the second harmonic coefficient $c_2$. The ratio over the average radius $c_2/c_0$ is the path length fraction of the maximal/minimal path against the average. Example values of $c_2/c_0$ for different collision systems and path length functions $R(\phi)$ are given in Appendix \ref{appendix:data_supplement}.
    \item Take two copies of the \Deltapt\ shifted \pT\ spectrum fit. Shift one copy up toward higher \pT\ according to the proportion $\delta p_{\mathrm{T}} = \Deltapt\cdot c_2/c_0$, and another copy down by the same proportion. These spectra enclose the original shifted spectrum. The upshifted and downshifted spectra are labeled $\rmd N/\rmd \Delta p_{\mathrm{T}}^+$ and $\rmd N/\rmd \Delta p_{\mathrm{T}}^-$ respectively.
    \item Our model estimate for the high-\pT\ differential \vtwo\ is then the difference weighted to the original spectrum:
    \begin{equation}
        \vtwo(\pT) = \frac{\frac{\rmd N}{\rmd \Delta p_{\mathrm{T}}^+} - \frac{\rmd N}{\rmd \Delta p_{\mathrm{T}}^-}}{ 4\cdot \frac{\rmd N}{\rmd \Delta p_{\mathrm{T}}} }.
    \end{equation}
    This estimation is similar to that used in Ref.~\cite{Zigic:2018ovr} extrapolated under the assumption of uniform energy loss $\Deltapt$, or can alternatively be thought of as a relation between the Fourier series coefficients of path length $r(\phi)$ and flow $v_n$ coefficients~\eqref{eq:azimuthal_anisotropy_fourier_decomp}.
\end{enumerate}
These steps are illustrated diagrammatically in Fig.~\ref{fig:process_v2}. The \vtwo\ offers a new arena to compare the classes \Ainc\ and \Awid\ observed in the transverse area. We will use the Full-Width-at-Half-Max contour from the width class \Awid\ and the average energy radius curve from the inclusive class \Ainc; the data for Glauber $c_2/c_0$ for these methods are shown in Table~\ref{tab:B:glauber_shape_harmonics_combined} of Appendix~\ref{appendix:data_supplement}.
\par Note that this model contains an assumption of linear path-length dependent energy loss $\Deltapt(L) \propto L$ to assert that the path length proportion $c_2/c_0$ can be translated directly to an energy loss for extra path-length $\delta p_{\mathrm{T}} = \Deltapt\cdot c_2/c_0$. Modeling a different dependence would require a more complicated relationship that reflects the nonlinear dependence on azimuthal angle: a different power would cause a mixing of Fourier components and other complications. We use the simple linear dependence for now, and relegate other powers to future study.

\section{Results} \label{sec:Results}

\begin{figure}[t]
    \centering
    \includegraphics[width=0.486\textwidth]{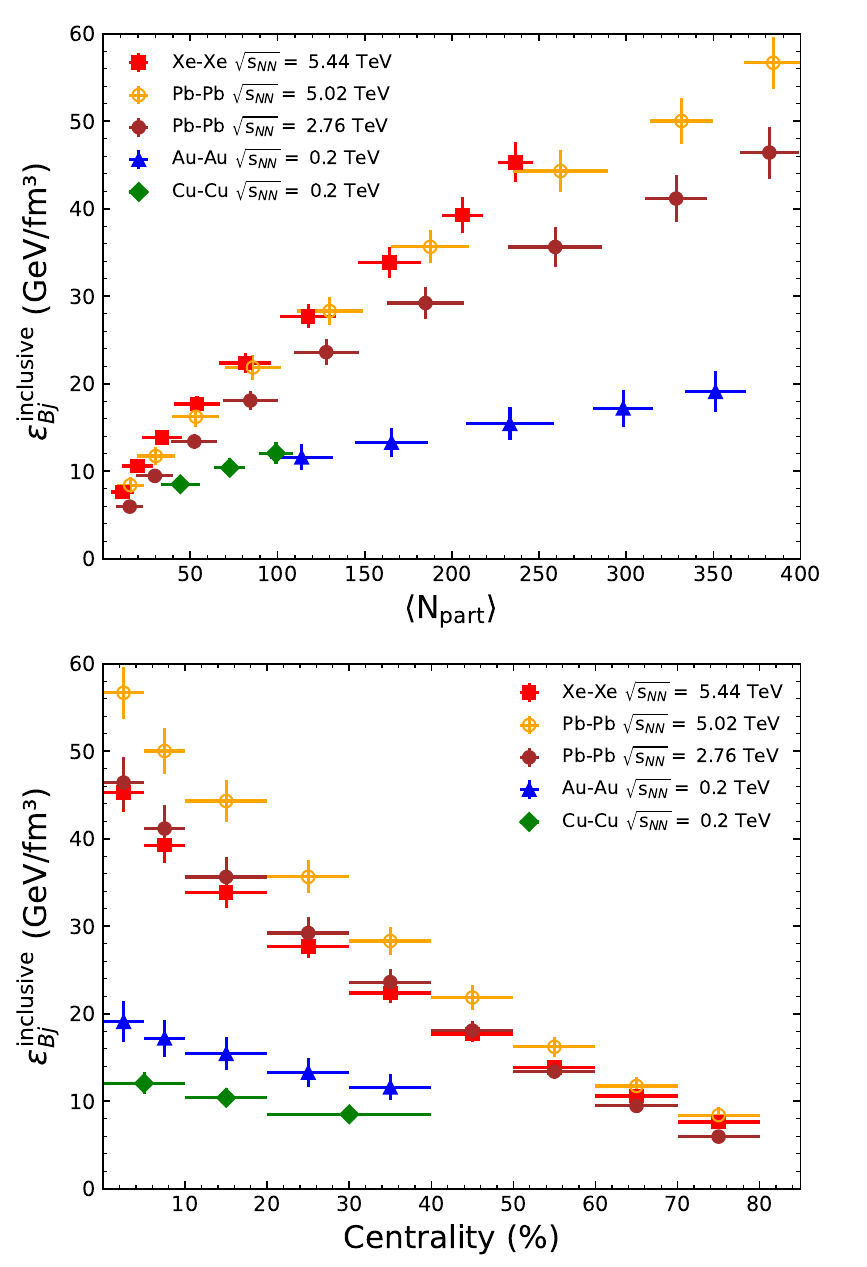} 
    \caption{Energy density using \Ainc\ for a variety of collision species and beam energies, as a function of $\langle N_{\text{part}} \rangle$ (upper) and centrality (lower).}
     \label{fig:ED-inc} 
\end{figure}

\begin{figure}[t!]
    \centering
    \includegraphics[width=0.5\textwidth]{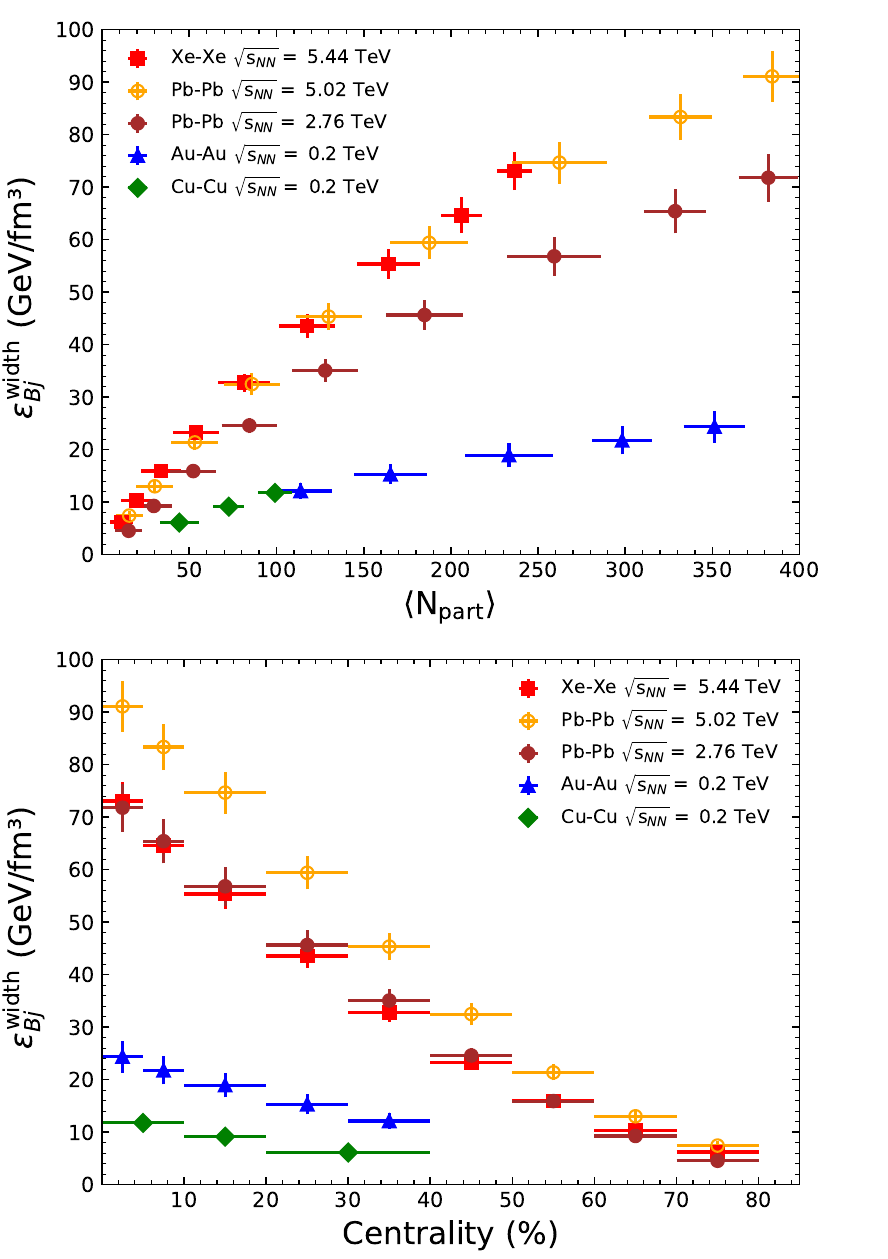} 
    \caption{Energy density using \Awid\ for a variety of collision species and beam energies, as a function of $\langle N_{\text{part}} \rangle$ (upper) and centrality (lower).} 
    \label{fig:ED-width} 
\end{figure}

Figures~\ref{fig:ED-inc} and~\ref{fig:ED-width} present the Bjorken energy densities, \ebjw\ and \ebji, respectively, as functions of centrality and \Npart\ for a variety of collision energies and species. Sensibly, \ebj\ increases monotonically with increasing \Npart\ or \sNN\ for both \ebjw\ and \ebji, but this does not generally hold for \ebje, as will be discussed later. While the energy densities are similar for the most peripheral data, the width based \ebjw\ rises more steeply with centrality, resulting in an approximately 50\% larger energy density estimate for the most central \PbPb\ data relative to \ebji. 

\begin{table}[b]
  \centering
  \renewcommand{\arraystretch}{1.5}
  \resizebox{\columnwidth}{!}{%
    \begin{tabular}{|l|c|c|c|c|}
      \hline
      Experiment ($\sqrt{s}$) & $C$ ($c$/GeV) & $n$ & $T$ (GeV) & \makecell{Reduced \\ MSE Metric} \\ 
      \hline
      ALICE (5.44 TeV) & 18.540 & 5.566 & 0.214 & 0.002423 \\ 
      \hline
      ATLAS (5.44 TeV)  & 18.540 & 5.555 & 0.209 & 0.001209 \\ 
      \hline
      ALICE (5.02 TeV)  &  9.825 & 5.684 & 0.247 & 0.003396 \\ 
      \hline
      ALICE (2.76 TeV)   & 10.720 & 5.970 & 0.227 & 0.004184 \\ 
      \hline
      STAR (0.2 TeV)     & 16.578 & 8.165 & 0.154 & 0.01000  \\ 
      \hline
      PHENIX (0.2 TeV)   & 16.578 & 9.108 & 0.169 & 0.00155   \\ 
      \hline
      STAR (0.2 TeV, $\pi^0$) & 95.266 & 8.169 & 0.112 & 0.000677 \\ 
      \hline
    \end{tabular}%
  }
  \caption{\pp\ cross section \pT\ spectra fit parameters ($C$, $n$, $T$) and the reduced MSE/DoF metric for the Tsallis fit to each dataset used in the analysis.}
  \label{table:ppMSE}
\end{table}

\begin{figure*}[p]
    \centering
    \includegraphics[width=0.90\textwidth]{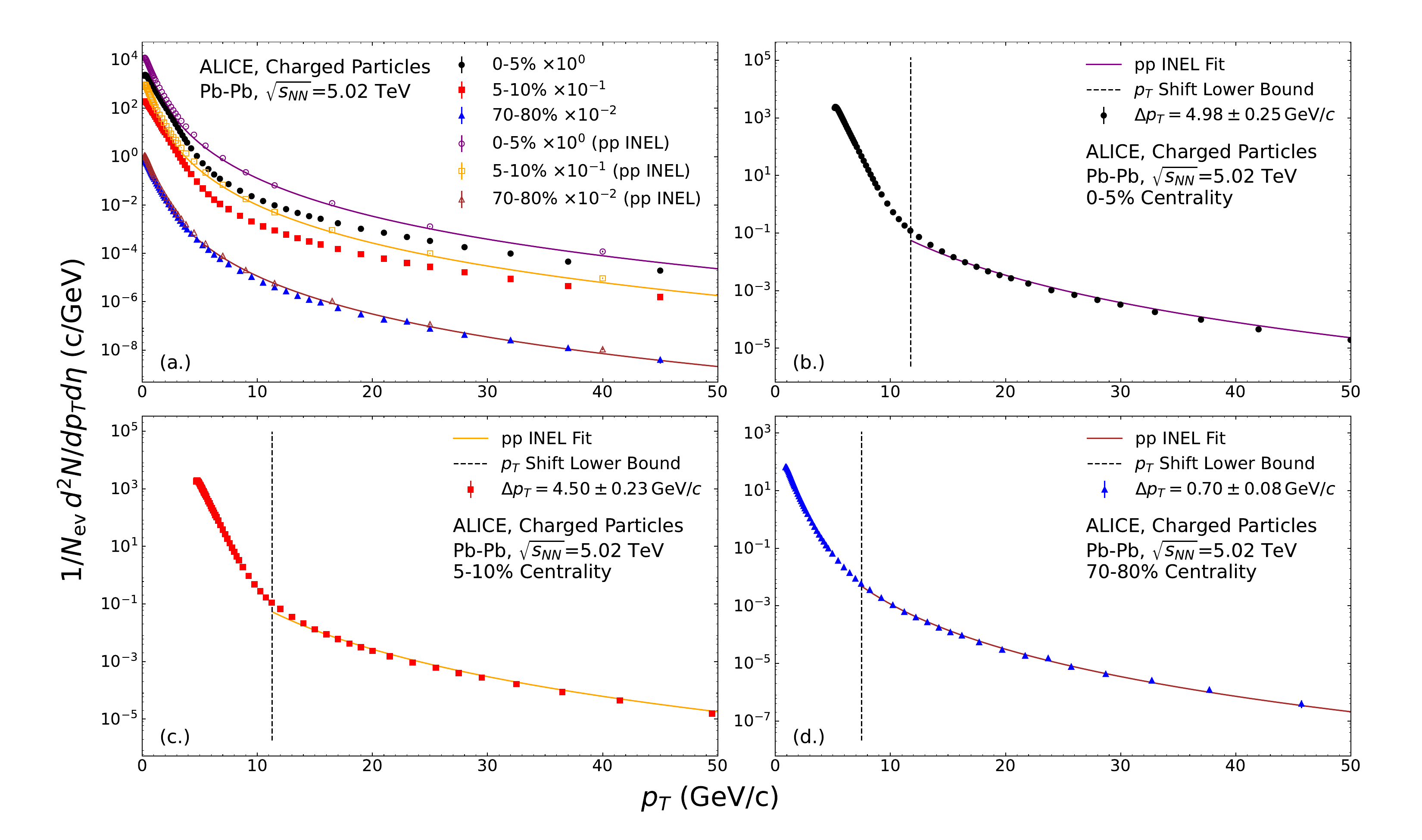} 
    \caption{\pT\ spectra of charged particles measured by the ALICE collaboration in \PbPb\ collisions at \sNN\ = 5.02 TeV for different centralities. The first panel shows the \pp\ Tsallis fit appropriately \Ncoll-scaled to each \PbPb\ dataset. The remaining three panels show the \Ncoll-scaled \pp\ Tsallis fit compared to the individual \Deltapt\ shifted \PbPb\ \pT\ spectrum in the region beyond the \pT\ shift lower bound. Uncertainty bars are present, but small relative to marker sizes.}
    \label{fig:PbSpectra}
\end{figure*}

\begin{figure*}[p]
    \centering
    \includegraphics[width=0.90\textwidth]{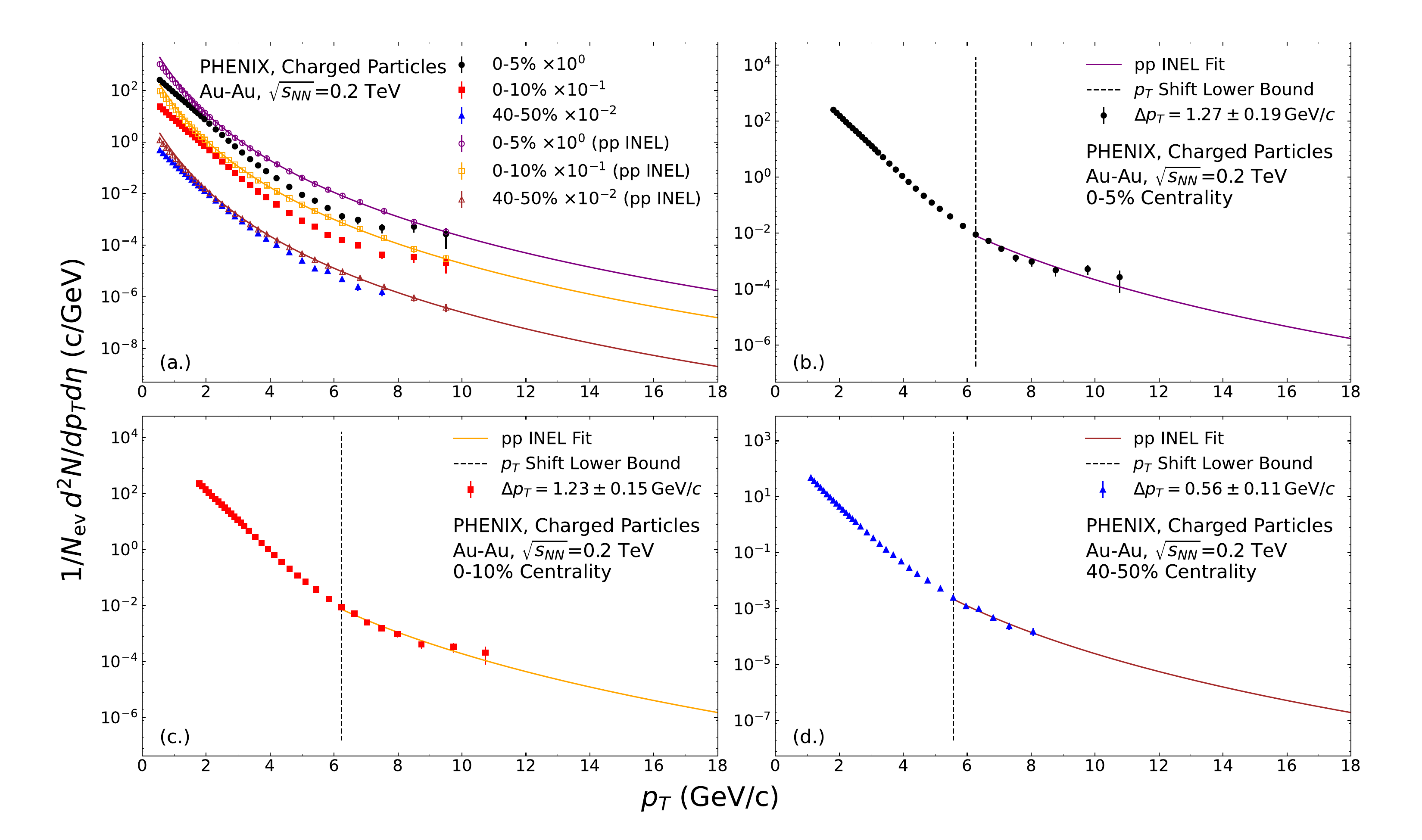} 
    \caption{ \pT\ spectra of charged particles measured by the PHENIX collaboration in \AuAu\ collisions at \sNN\ = 200 GeV for different centralities. The first panel shows the \pp\ Tsallis fit appropriately \Ncoll-scaled to each \AuAu\ dataset. The remaining three panels show the \Ncoll-scaled \pp\ Tsallis fit compared to the individual \Deltapt\ shifted \AuAu\ \pT\ spectrum in the region beyond the \pT\ shift lower bound.}
    \label{fig:AuPHENIX}
\end{figure*}

\begin{figure*}[p]
    \centering
    \includegraphics[width=0.9\textwidth]{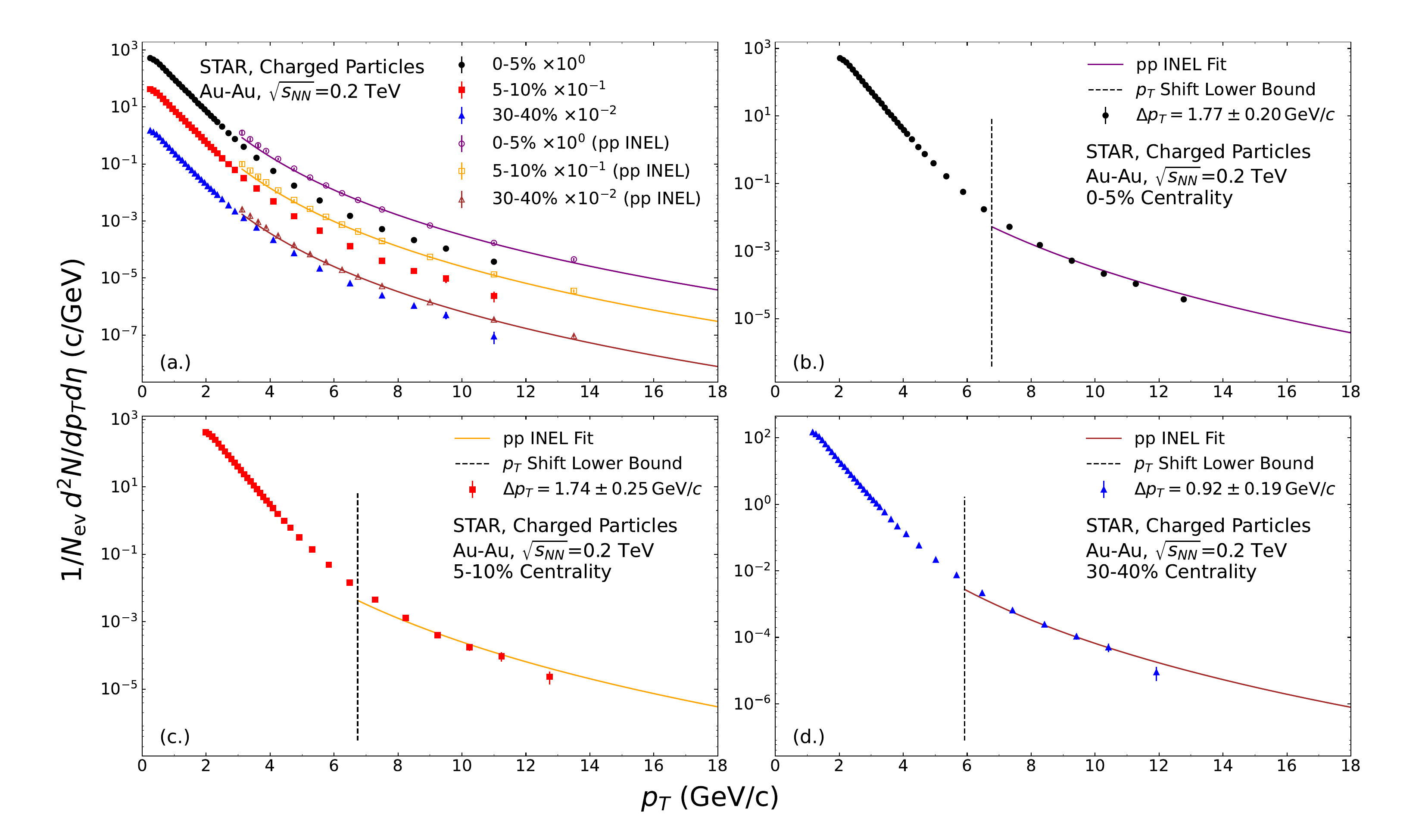} 
    \caption{ \pT\ spectra of charged particles measured by the STAR collaboration in \AuAu\ collisions at \sNN\ = 200 GeV for different centralities. The first panel shows the \pp\ Tsallis fit appropriately \Ncoll-scaled to each \AuAu\ dataset. The remaining three panels show the \Ncoll-scaled \pp\ Tsallis fit compared to the individual \Deltapt\ shifted \AuAu\ \pT\ spectrum in the region beyond the \pT\ shift lower bound.}
    \label{fig:AuSTAR}
\end{figure*}

Following the procedure in Sec.~\ref{subsec:Deltapt}, illustrative \Deltapt\ results from ALICE \PbPb\ at \sNN\ = 5.02 TeV as well as PHENIX and STAR \AuAu\ at \sNN\ = 200 GeV datasets are shown in Figs.~\ref{fig:PbSpectra},~\ref{fig:AuPHENIX} and~\ref{fig:AuSTAR} respectively. In each figure, the top left panel (a.) shows the charged particle \pT\ spectra for three sample \AA\ centrality bins alongside the corresponding appropriately \Ncoll-scaled \pp\ inelastic data. The curves are the \Ncoll-scaled Tsallis fits to the \pp\ reference spectra. The fit parameters and MSE metric for all resultant \pp\ data fits are provided in Table~\ref{table:ppMSE}. The suppression of the \AA\ spectra with respect to the \TAA-scaled Tsallis \pp\ fit is evident in all cases, in agreement with published \RAA\ values~\cite{ALICE:2018vuu, ALICE:2018hza,ATLAS:2022kqu,STAR:2003fka,PHENIX:2003djd,STAR:2009mwg}. The remaining three panels (b.\textendash d.) show the \Deltapt-shifted \AA\ data for the three sample centrality bins alongside the Tsallis \pp\ fit curve. The \AA\ spectra have been shifted by the \Deltapt\ which best aligns the two and minimizes the MSE metric above the determined momentum threshold \pTmin\ denoted by the vertical line. Note that the \pTmin\ line has, along with the \AA\ data points, been shifted rightward by \Deltapt\ in panels (b.\textendash d.). The  Tsallis \pp\ fit curve is only shown in the relevant region above this \pTmin\ threshold.  The strong visual agreement between the shifted \AA\ spectra and \Ncoll-scaled \pp\ reference Tsallis fit across the wide range of species, energy, and centrality further validates the MSE as a fit metric, and affirms the approach explored in this study in which a single energy loss \Deltapt\ is determined for each centrality bin. Note that the final MSE metric for each \Deltapt\ fit, including those shown in Figs.~\ref{fig:PbSpectra},~\ref{fig:AuPHENIX} and~\ref{fig:AuSTAR} is provided for each collision system and centrality bin in Appendix \ref{appendix:data_supplement}. As a further model check, the extracted \Deltapt-shifted Tsallis fits were substituted for the \AA\ data, and ratios against the corresponding \pp\ baseline spectra were taken for comparison against published \RAA\ data. Rough agreement of this model ratio with \RAA\ was observed within $2\sigma$ for all collision systems studied, though it should be noted that the model deviates more strongly from \RAA\ in central collisions. 

\begin{table}[t]
    \centering
    \renewcommand{\arraystretch}{0.95}
    \resizebox{\columnwidth}{!}{%
    \begin{tabular}{|c|c|c|c|c|}
    \hline
    \makecell{Centrality \\ Bin (\%)} 
    & \multicolumn{2}{c|}{\makecell{STAR $pp$ reference \\ $\sNN\ = 200$ GeV}} 
    & \multicolumn{2}{c|}{\makecell{PHENIX $pp$ reference\\ $\sNN = 200$ GeV}} \\
    \hline
    & STAR & PHENIX  & STAR  & PHENIX  \\
    & \AuAu\  \Deltapt & \AuAu\  \Deltapt & \AuAu\  \Deltapt & \AuAu\  \Deltapt \\
    & (GeV/$c$) & (GeV/$c$) & (GeV/$c$) & (GeV/$c$) \\
    \hline
    0–5\%   & $1.77\pm0.23$ & $1.55\pm0.22$ 
            & $1.38\pm0.13$ & $1.27\pm0.21$ \\
    10–20\% & $1.43\pm0.22$ & $1.40\pm0.22$ 
            & $1.07\pm0.16$ & $1.12\pm0.21$ \\
    20–30\% & $1.18\pm0.20$ & $1.04\pm0.21$ 
            & $0.85\pm0.15$ & $0.81\pm0.20$ \\
    30–40\% & $0.92\pm0.20$ & $0.62\pm0.09$ 
            & $0.62\pm0.17$ & $0.63\pm0.12$ \\
    40–50\% & -             & $0.72\pm0.13$ 
            & -             & $0.56\pm0.12$ \\
    \hline
    \end{tabular}
    }
    \caption{\Deltapt\ shift results for STAR and PHENIX $\sNN = 200$ GeV \AuAu\ spectra, using either the PHENIX or STAR \pp\ charged particle cross section as a common reference.}
    \label{table:star-phenix}
\end{table}

\begin{table*}[p]
    \centering
    \renewcommand{\arraystretch}{1.4}
    \begin{tabular}{|c|c|c|c|c|c|c|c|}
        \hline
        Experiment & Beam Species, \sNN 
        & \makecell{\ebjw\ Intercept\\ (GeV/c)}  & \makecell{\ebjw\ slope\\ (fm$^3$/c)}
        & \makecell{\ebji\ Intercept\\ (GeV/c)} & \makecell{\ebji\ slope\\ (fm$^3$/c)}
        & \makecell{Reduced \\ $\chi^2_{\mathrm{width}}$}
        & \makecell{Reduced \\ $\chi^2_{\mathrm{inclusive}}$}\\
        \hline
        ALICE   & \XeXe, 5.44 TeV & $-0.164 \pm 0.069$ & $0.057 \pm 0.002$ & $-0.689 \pm 0.121$ & $0.106 \pm 0.006$ & $0.50$ & $0.96$ \\
        ATLAS   & \XeXe, 5.44 TeV & $-0.170 \pm 0.049$ & $0.054 \pm 0.002$ & $-0.669 \pm 0.086$ & $0.100 \pm 0.004$ & $0.17$ & $0.33$ \\
        ALICE   & \PbPb, 5.02 TeV & $0.295 \pm 0.038$ & $0.049 \pm 0.001$ & $-0.095 \pm 0.025$ & $0.090 \pm 0.001$ & $0.23$ & $0.06$ \\
        ALICE   & \PbPb, 2.76 TeV & $0.466 \pm 0.067$ & $0.049 \pm 0.002$ & $0.144 \pm 0.081$ & $0.086 \pm 0.004$ & $0.38$ & $0.36$ \\
        STAR    & \AuAu, 200 GeV   & $0.027 \pm 0.103$ & $0.074 \pm 0.006$ & $-0.489 \pm 0.188$ & $0.124 \pm 0.013$ & $0.07$ & $0.12$ \\
        PHENIX  & \AuAu, 200 GeV   & $0.035 \pm 0.110$ & $0.055 \pm 0.007$ & $-0.410 \pm 0.203$ & $0.096 \pm 0.015$ & $0.34$ & $0.50$ \\
        STAR    & \CuCu, 200 GeV   & $-0.131 \pm 0.078$ & $0.063 \pm 0.009$ & $-0.603 \pm 0.142$ & $0.101 \pm 0.014$ & $0.17$ & $0.16$ \\
        Global  &                   & $0.047 \pm 0.046$ & $0.054 \pm 0.002$ & $-0.429 \pm 0.066$ & $0.099 \pm 0.003$ & $1.47$ & $1.78$ \\
        \hline
    \end{tabular}
    \caption{Linear fit results for \Deltapt\ as a function of \ebjw\ and \ebji. Fitting is performed using orthogonal distance regression to incorporate uncertainties in dependent and independent axes. Reduced $\chi^2$ computed using only independent axes uncertainties are included.}
    \label{table:fit_results}
\end{table*}

\par It should be noted that the Tsallis power law parameters $n$ fit to STAR and PHENIX \pp\ data at $\sqrt{s} = 200$ GeV differ significantly, while all fit parameters for the ALICE and ATLAS data at $\sqrt{s} = 5.44$ TeV are consistent. The disagreement in the RHIC fit results is likely due differences in how the charged particle cross section was determined by the two collaborations. PHENIX computed charged particle \pT\ spectra by extrapolating the $\pi^0$ spectrum from Ref.~\cite{PHENIX:2003fvg}. STAR, on the other hand, determined the \pp\ charged particle \pT\ spectra by summing measured \pT-differential $\pi^{\pm}$, $K^{\pm}$, $p$ and $\overline{p}$ yields~\cite{STAR:2011iap}. The STAR determination results in a more significant flattening of the higher \pT\ data compared to the PHENIX extrapolation, reflected in the smaller power law $n$.

Following the discrepancy in \pp\ reference spectra, the \Deltapt\ values calculated for \AuAu\ collisions at $\sNN = 200$ GeV differ between the STAR and PHENIX datasets. This is again in contrast to the \XeXe\ data reported by ATLAS and ALICE, where we see agreement within uncertainties for resultant \Deltapt\ values across all centrality bins. To examine whether this discrepancy in \Deltapt\ is purely due to the \pp\ spectra differences observed in the fits, we compared \Deltapt\ values obtained from STAR and PHENIX \AuAu\ spectra with a common fixed \pp\ reference spectrum. Table~\ref{table:star-phenix} shows the \Deltapt\ calculated when either the STAR or PHENIX \pp\ charged particle spectrum is used as the common \pp\ reference. In both cases, the computed \Deltapt\ values for each collaboration's \AuAu\ data agree within uncertainties. This affirms that the observed differences in \Deltapt\ is dominantly an artifact of the differences in \pp\ spectral shape between collaborations. For consistency, however, our analysis proceeds using the \pp\ spectrum reported by each collaboration to calculate the \Deltapt\ for their respective \AA\ data in our final results.

\par
\begin{figure}[h!]
    \centering
    \includegraphics[width=0.495\textwidth]{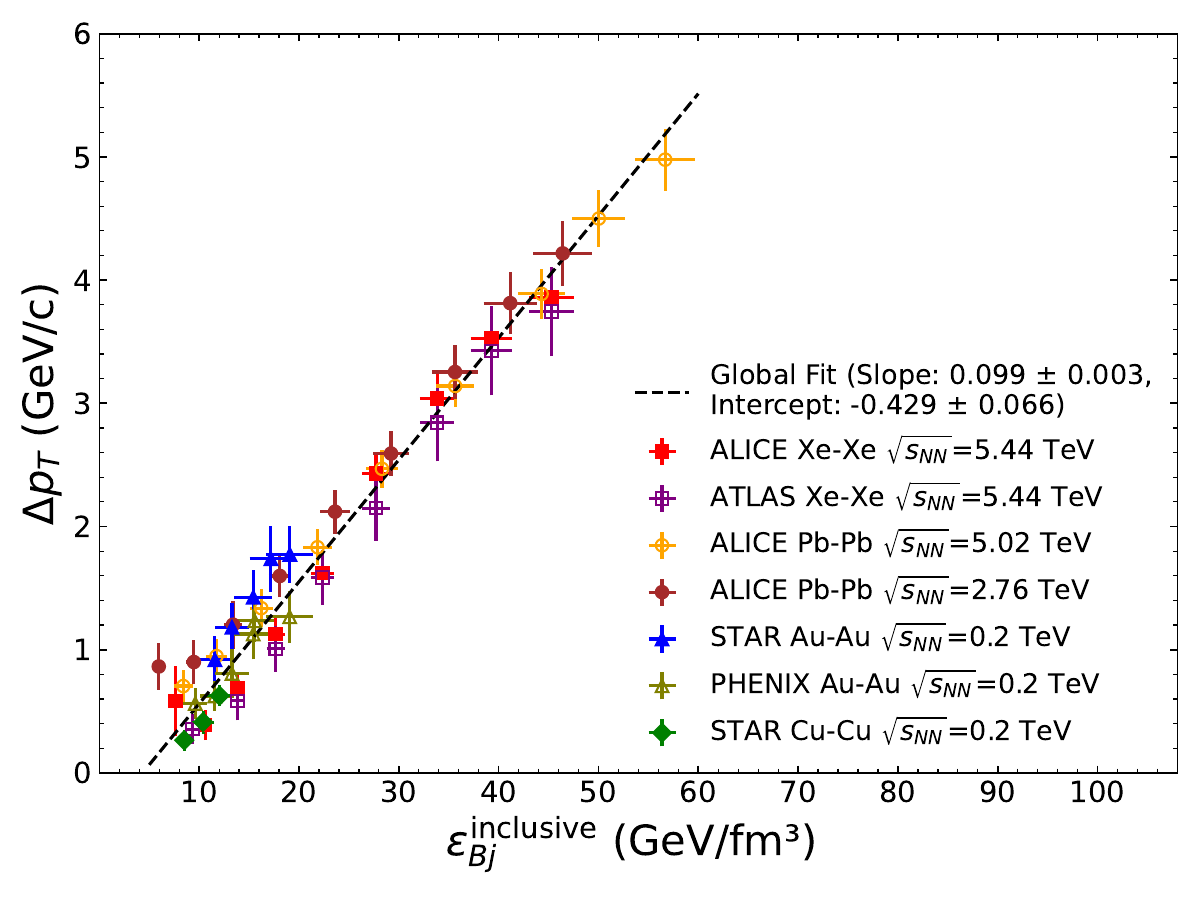} 
    \includegraphics[width=0.495\textwidth]{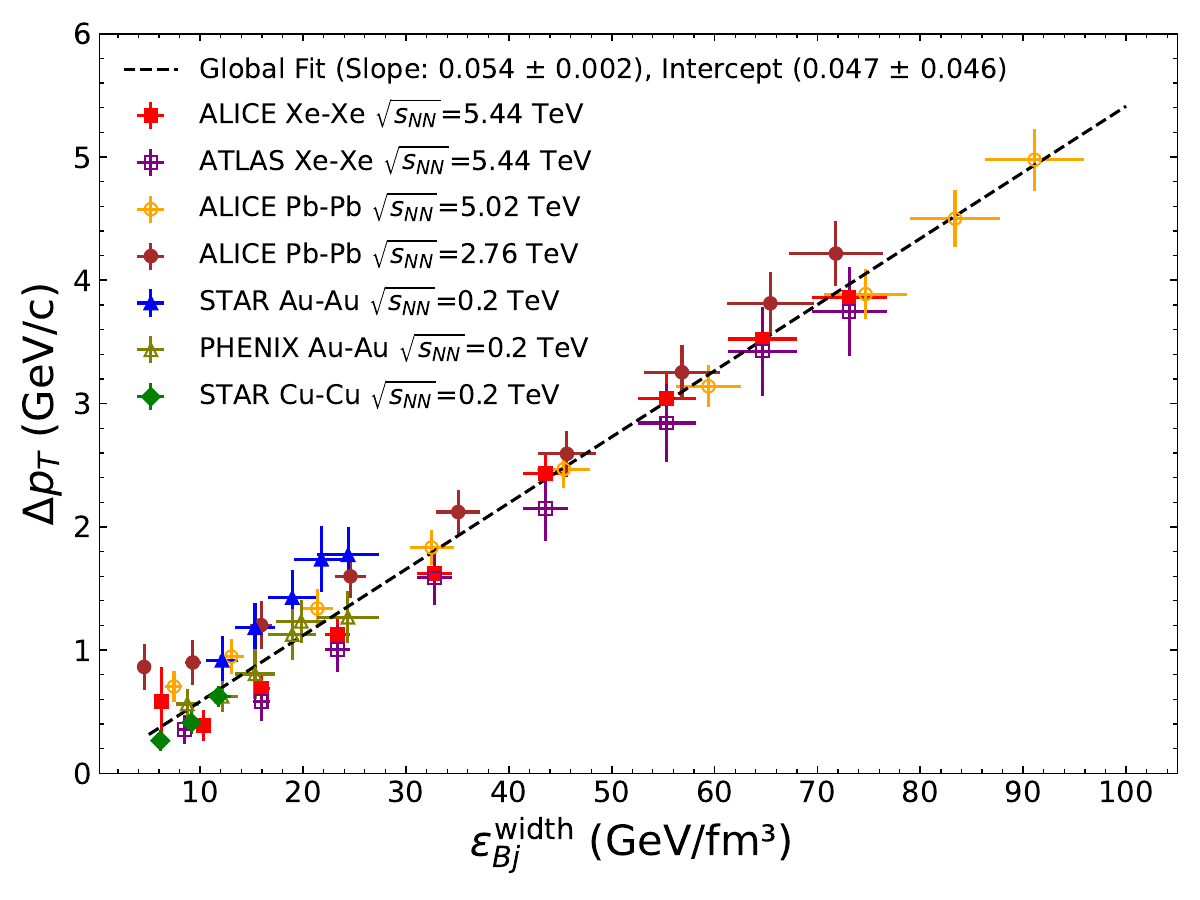} 
    \includegraphics[width=0.495\textwidth]{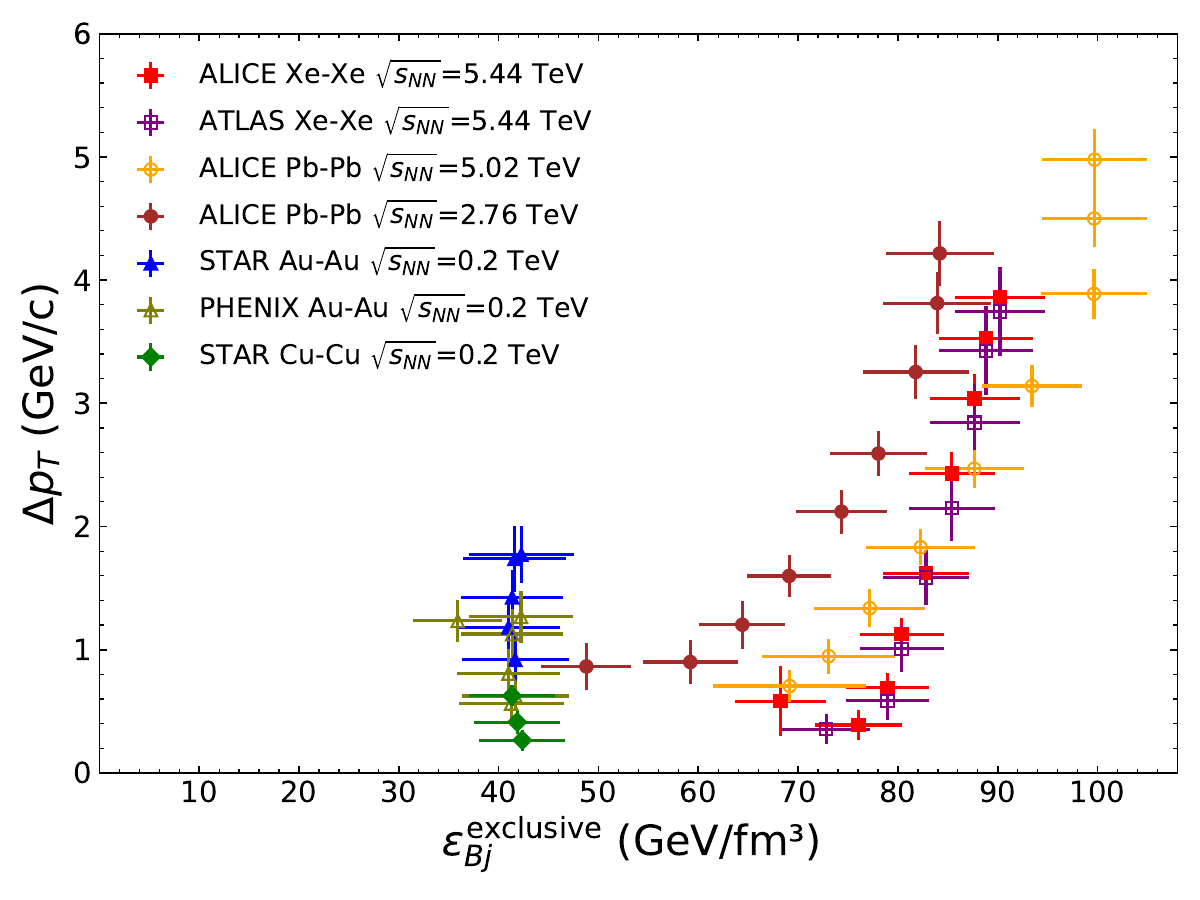} 
    \caption{Energy density \ebj\ versus \Deltapt\ for a variety of collision species and beam energies. Estimates of energy density from each area scaling class \Ainc, \Awid, and \Aexc\ are shown separately. Note the STAR \CuCu\ data use reported $\pi^\pm$ spectra for calculating \Deltapt\ while  other datasets use inclusive charged particle spectra.} 
    \label{fig:width-fit} 
\end{figure}

 With this understanding of \Deltapt\ and \ebj\ in hand, we can now move on to the central objective of this work: characterization of the relationship between these two quantities. The resulting \Deltapt\ as functions of Bjorken energy density for each area class are shown in Fig.~\ref{fig:width-fit}. For the inclusive and width-based areas, a clear direct correlation of the high-\pT\ charged particle \Deltapt\ with the estimated initial energy density \ebj\ is observed, from peripheral \CuCu\ collisions at $\sNN\ = 200$ GeV to central \PbPb\ events at $\sNN\ = 5.02$ TeV. To evaluate whether these correlations are linear to leading order, a $\chi^2$ fit to a variable power $\Deltapt = A(\ebj / \epsilon_0)^{p} + b$ is performed, from which the resulting powers are $p^{\rm inclusive} = 1.006^{+0.115}_{-0.112}$ and $p^{\rm width} = 1.016^{+0.083}_{-0.080}$. These correspond to $\chi^2 / \mathrm{DoF}$ of 58.77/46 and 57.56/46 for the inclusive and width-based methods respectively, strongly supporting the conclusion of a linear relationship at first order.

Performing linear regressions, a slope of 0.054 $\pm$ 0.002 $\text{fm}^3/c$ and 0.099 $\pm$ 0.003 $\text{fm}^3/c$ for the width and inclusive area estimates respectively. The individual linear fits for each experiment-collision pair, as well as the corresponding reduced $\chi^2$ values are provided in Table \ref{table:fit_results}. Such a strong correlation indicates that the initial energy density of the overlap region primarily drives partonic energy loss, and that other factors such as the shape of the overlap region are of secondary importance at best. In addition, this correlation is preserved through the fragmentation and hadronization process such that it can be detected in the single hadron \pT\ spectra. It also appears that the manner of energy density generation is of little consequence; collisions of low $A$ species at high energy or high $A$ species at low energy result in the same \Deltapt, so long as QGP is formed and events are selected with equivalent initial energy densities. To explore further for evidence of subleading dependences, the residuals $\delta(\Deltapt)$ to the linear fits shown in Fig.~\ref{fig:width-fit} are studied as a function of effective path length $\L_{\rm eff}$ and path difference $\Delta L_{\rm eff}$ as well as \Npart\ in Appendix~\ref{appendix:deltapt_residuals}.

\par Although \Aexc\ presents itself as an outlier in Fig. \ref{fig:areascale_PbPb2.76}, we include the relationship between \Deltapt\ and \ebje\ in Fig.~\ref{fig:width-fit} for completeness. We see that for collisions at RHIC energies, the calculated energy density using \Aexc\ is approximately constant for all data points, indicating that the ratio of \dnchdeta\ and \Aexc\ does not vary over centrality or species. In the specific case of STAR \CuCu, \ebje\ slightly decreases with increasing impact parameter, suggesting nonphysically that peripheral collisions produce larger energy density than central collisions at the same \sNN. For LHC data however, \ebje\ does rise with centrality and \Deltapt\ roughly increases with energy density, but these behaviors are different for each collision system. While in principle \Aexc\ could provide a reasonable estimate of transverse area, our observations of the inconsistent scaling between RHIC and LHC data, the disagreement of centrality scaling with all other area methods shown in Fig.~\ref{fig:areascale_PbPb2.76} and Fig.~\ref{fig:appA:all_areas}, and the previously reported strange behaviors in ALICE \pPb~\cite{ALICE:2022imr} culminate to strongly disfavor this area scaling for the computation of \ebj. 

\par
\begin{figure}[t]
    \centering
    \includegraphics[width=0.98\linewidth]{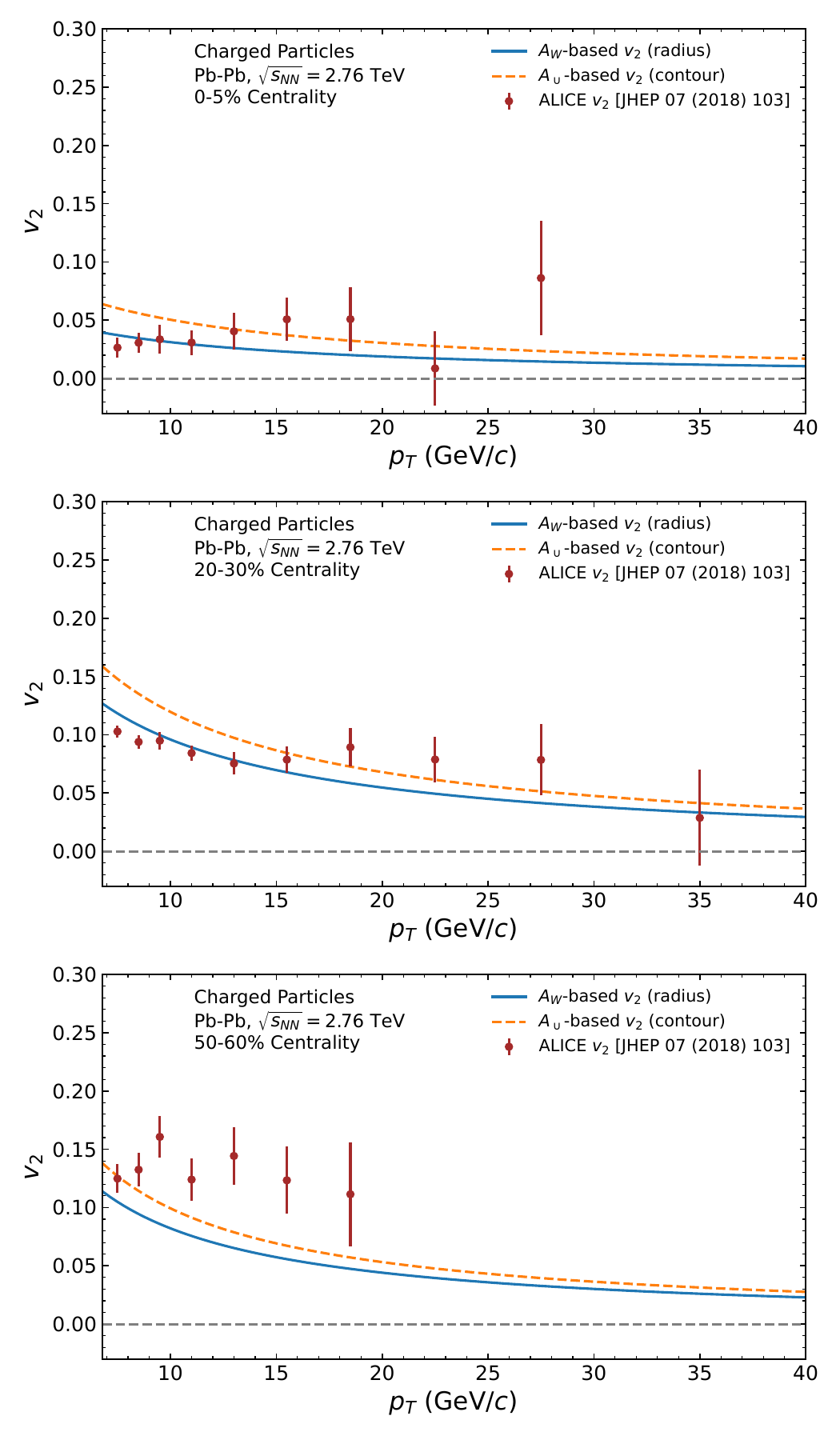}
    \caption{High-\pT\ $v_2$ estimations derived from \Deltapt\ as well as the avg. energy radius/avg. pressure radius area definitions, corresponding to the \Awid\ class (solid blue curve) and the FWHM contour area definition, corresponding to the \Ainc area class (dashed orange curve), compared to ALICE experimental data from \PbPb\ collisions at \sNN\ = 2.76 TeV.}
    \label{fig:v2-PbPb}
\end{figure}

\begin{figure}[t]
    \centering
    \includegraphics[width=0.98\linewidth]{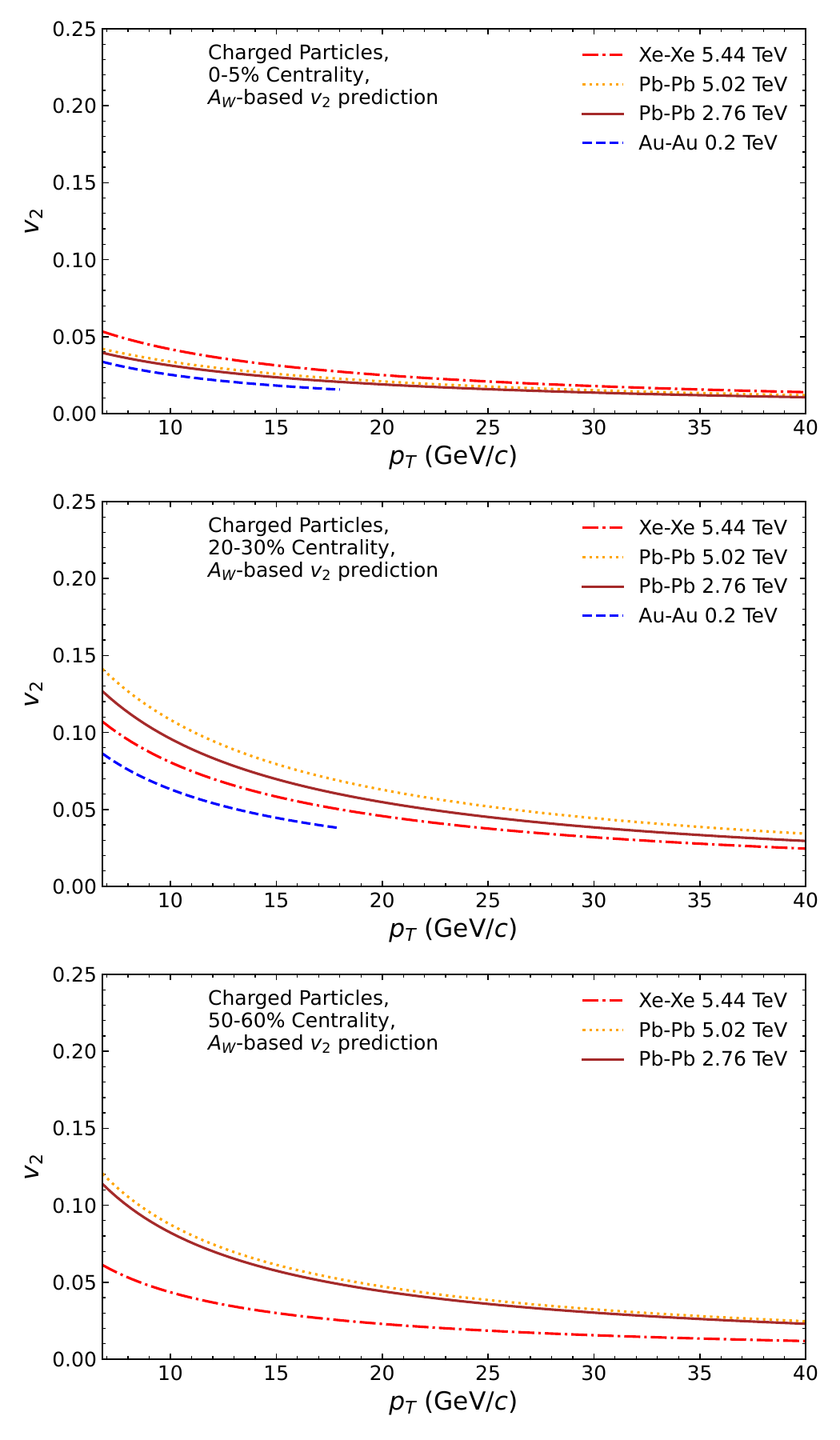}
    \caption{High \pT\ $v_2$ predictions using the \Deltapt\ and $A_W$ based model for a variety of collision systems.}
    \label{fig:v2-OtherCollisions}
\end{figure}

 Continuing to \vtwo\ results, Fig.~\ref{fig:v2-PbPb} shows the prediction for $\sNN = 2.76$ TeV \PbPb\ high-\pT\ \vtwo\ using the procedure described in Sec. \ref{subsec:v2}. Model predictions for the \Awid\ and \Ainc\ classes are compared to corresponding ALICE \vtwo\ data~\cite{ALICE:2018rtz} for three different centrality bins. Predictions for both area estimates are similar in shape but differ slightly in normalization, with the \Awid\ class favored by central data and \Ainc\ performing better for peripheral collisions. Considering the simplicity of the modeling, a reasonable agreement is obtained across the available \pT\ range, within $3\sigma$ for all data for \Awid-based \vtwo. Stronger precision on high-\pT\ \vtwo\ data could better distinguish between the two models or clarify regimes of model failure, such as in peripheral collisions where there is more tension between the models and data.
\FloatBarrier 
Encouraged by the general agreement with the magnitude of \vtwo\ across the \pT\ range shown in ALICE \PbPb\ data, Fig.~\ref{fig:v2-OtherCollisions} presents our model's predictions for the other collision systems studied. As it performs slightly better for the ALICE \PbPb, the predictions utilize the initial overlap regions' harmonics $c_2,\,c_0$ derived from the width-based area estimates, though the corresponding \vtwo\ for \Ainc\ are shown in Appendix~\ref{Appendix:v2_inclusive}. The \vtwo\ for the \AuAu\ data at \sNN\ = 200 GeV is the lowest at a fixed \pT\ and centrality; although the $c_2/c_0$ ratios in \AuAu\ are similar to those for the LHC \PbPb\ and \XeXe, the extracted \Deltapt\ are significantly reduced due to the smaller \sNN, resulting in smaller \vtwo\ in our modeling. 

The trends of the \XeXe\ predictions are especially interesting, as the \sNN\ = 5.44 TeV \XeXe\ contains the largest \vtwo\ among all species for the most central data, which falls to the smallest LHC \vtwo\ in peripheral collisions. A similar flipping of the \XeXe\ relative to \sNN\ = 5.02 TeV \PbPb\ is also predicted by hydrodynamic calculations~\cite{Giacalone:2017dud} and observed in data~\cite{ALICE:2018lao, ATLAS:2019dct}, albeit at significantly lower \pT. The enhancement of low-\pT\ \vtwo\ in central \XeXe\ is attributed primarily to nuclear shape, as the quadrupole deformation $\beta_2$ is expected to be much larger in \ce{^129Xe} than \ce{^208Pb}~\cite{Moller:2015fba}. The nuclear size also plays a role; systematic analyses of $v_n$ observables have shown that statistical fluctuations in yields produce a universal enhancement of all $v_n$ in systems with fewer sources. The enhancement of $v_3$ in \XeXe\ relative to \PbPb\ across the centrality range is considered a signal of this effect~\cite{ALICE:2018lao}. 

The depletion of \vtwo\ for peripheral \sNN\ = 5.44 TeV \XeXe\ collisions relative to \sNN\ = 5.02 TeV \PbPb\ is therefore unexpected on the basis of nuclear structure alone, as models indicate that the nuclear shape of \ce{^129Xe} has negligible effects in the peripheral region~\cite{Giacalone:2017dud}. This signal must therefore be the result of differing medium properties. In the case of the low-\pT\ measurements by ALICE~\cite{ALICE:2018lao} and ATLAS~\cite{ATLAS:2019dct}, the depletion is attributed to nonzero shear viscosity $\eta/s$ during hydrodynamic evolution; hydrodynamic calculations~\cite{Giacalone:2017dud} reproduce the depletion only for nonzero $\eta/s$, though calculations are similar for a range of $\eta/s$~\cite{ALICE:2018lao}. In the absence of significant hydrodynamic effects for high-\pT\ hadrons, the depletion is instead attributed to path-length dependent jet quenching~\cite{Harris:2024aov}. 
While the \pT\ ranges reported in Refs.~\cite{ALICE:2018lao,ATLAS:2019dct} are restricted due to the limited statistics from the short \XeXe\ run, future precision measurements at the LHC with a range of nuclear species could be very impactful to our understanding of the path length dependence of partonic energy loss.
  
\section{Conclusions} \label{sec:Conc}

In summary, we have demonstrated a strong linear correlation between the estimated initial energy density \ebj\ created in a heavy-ion collision and the average energy loss \Deltapt\ of high-\pT\ charged particles generated by those collisions. This striking correlation is observed across a wide variety of collision systems, from \AuAu\ and \CuCu\ collisions at \sNN\ = 200 GeV to \PbPb\ and \XeXe\ events at $\sNN > 5$ TeV. The observed correlation appears independent of ion species and collision energy, and persists among disparate methods of Glauber modeling. This result suggests that the initial energy density is the driving factor in predicting the average energy loss of hard scattered partons to the QGP. Detailed knowledge of other factors such as the initial event eccentricity and parton composition are not critical for predicting the azimuthally averaged charged particle \RAA.

A variety of methods were studied to extract the transverse overlap area using Monte-Carlo Glauber calculations, needed to compute \ebj. These methods included three grid-based EBE calculations described in Ref.~\cite{Loizides:2016djv}, along with five novel phenomenological methods. Two classes were identified based on the methods' approximate scalings with centrality, with the EBE \Aexc\ calculation being the sole outlier. The strange scaling for \Aexc\ we observe is also seen in Glauber estimates for ALICE \pPb~\cite{ALICE:2022imr}. Of the two identified classes, one scaled with centrality in a similar fashion to the EBE ``inclusive" \Ainc\ area, while the other class scaled with the ``width-based" \Awid\ EBE calculation. This classification of scaling persisted across all species and energies studied. Generally, the \Awid\ class displayed a stronger dependence on centrality than the \Ainc\ class, especially for more peripheral events. The \ebj\ estimated using \Aperp\ from either \Awid\ or \Ainc\ classes and the reported experimental \dnchdeta\ show the expected trends with centrality, \Npart, and collision energy, while the same is not true for the outlier \Aexc. Because of this, \Aexc\ does not produce the strong \Deltapt\textendash\ebj\ correlation observed for the other seven methods. The persistence of the correlation between both the \Ainc\ and \Awid\ area classes, with no change in \Deltapt\ modeling, supports that the correlation is robust and not due to Glauber model self-correlation between \Aperp\ inputs to \ebj\ and \Ncoll\ scaling of \Deltapt\ fit spectra. That the correlation then fails for \Aexc\ demonstrates that the correlation is not a trivial consequence of the model construction. In addition, the \Ncoll\ uncertainty estimation procedure, applied only to the computation of \Deltapt, does not significantly increase uncertainty or blur the correlation.

Good descriptions of the \AA\ high-\pT\ spectra can be obtained for all systems, collision energy, and centralities studied via \Deltapt\ shifting of the appropriately \Ncoll-scaled Tsallis function fits to the \pp\ \pT\ spectra. For simplicity a constant \pT\ shift is assumed for each spectrum. While theoretical arguments suggest a fractional energy loss with parton type and \pT\ dependence, the limited high-\pT\ charged particle range currently available combined with the smearing in \pT\ between parton initiator and final-state hadron appears to make this constant \Deltapt\ approximation reasonable. The effectiveness of this approximation over such a wide range of collision systems and energy may offer new insights into the deconvolution of medium-driven \pT\ spectrum modifications from kinematic ones.

We extended our model to explore predictions of high-\pT\ hadronic anisotropy \vtwo\ by coupling Glauber estimates of event geometry to observed \pT\ spectra in a novel approach. The model provides estimates of \vtwo\ that are reasonably consistent with ALICE data~\cite{ALICE:2018rtz} in the highest \pT\ data of each centrality bin, but overestimates \vtwo\ for lower \pT\ data. As the breakdown region of the model is still well above the collective region, the deviation of \vtwo\ from our model may indicate that our assumption of linear path length-dependent energy loss is ineffective for this mid-\pT\ region. We also observe a flipping of the magnitude of the \XeXe\ \vtwo\ at $\sNN = 5.44$ TeV relative to that in \PbPb\ at $\sNN = 5.02$ TeV when going from central to peripheral data. This observation indicates that our model is capable of accessing path-length dependent jet quenching and nuclear deformation through only initial state models and final-state \pT\ spectra. 

The interrelations between \Deltapt, \ebj, \vtwo\ and related quantities merit further research. While the exclusive area \Aexc\ exhibits strange behaviors when used for energy density computation, it may be valuable in other contexts. The division of the remaining area methods into two classes further complicates the question of whether a single \Aperp\ calculation exists, or what might separate these two classes. While our \pT-independent \Deltapt\ model serves as a reasonable approximation for interpreting high-\pT\ hadron spectra, it would be interesting to explore whether this approximation holds for jet spectra which are understood to serve as better, but not perfect, approximations of the initial parton's energy. Similarly, comparing our model's predictions with \pT-differential jet \vtwo\ may provide a more precise exploration of path-length dependent energy loss, especially if non-linear energy loss dependence is incorporated. Increased precision of jet measurements from LHC Run3, first results from sPHENIX and improved precision from upgraded STAR data should allow such detailed comparisons in the jet regime to be made soon. 

As the field begins to enter the precision era for both experimental measurements and theoretical modeling, simple models with few degrees of freedom such as this remain helpful for calibrating assumptions, (re)evaluating observables and highlighting the dominant necessary features that must be reproduced by more elaborate simulation frameworks and rigorous first-principle calculations.

\vspace{0.5cm}

\section*{ACKNOWLEDGMENTS}

We thank Constantin Loizides for helpful discussions about the interpretation of the Glauber Monte Carlo calculations, and Raymond Ehlers for insights about \pp\ \pT\ spectra fitting and fit functions. We gratefully acknowledge Christine Nattrass and Brant Johnson for their help in obtaining PHENIX \pp\ reference yields for analysis in this study. RH and HC are supported by DoE grant DE-SC004168.

\bibliography{MyBib}
\clearpage
\onecolumngrid


\appendix 
\section{Additional Information Concerning Energy Density, \Deltapt\ and Event Averaged Harmonic Calculations}
\label{appendix:data_supplement}

This appendix provides further details on the area calculations and other input data used to compute the energy density \ebj\ and the averaged event harmonics $c_2,\, c_0$ used in the elliptic flow \vtwo\ predictions.

\par
\begin{figure*}[h]
    \centering
    \includegraphics[height=0.92\textheight]{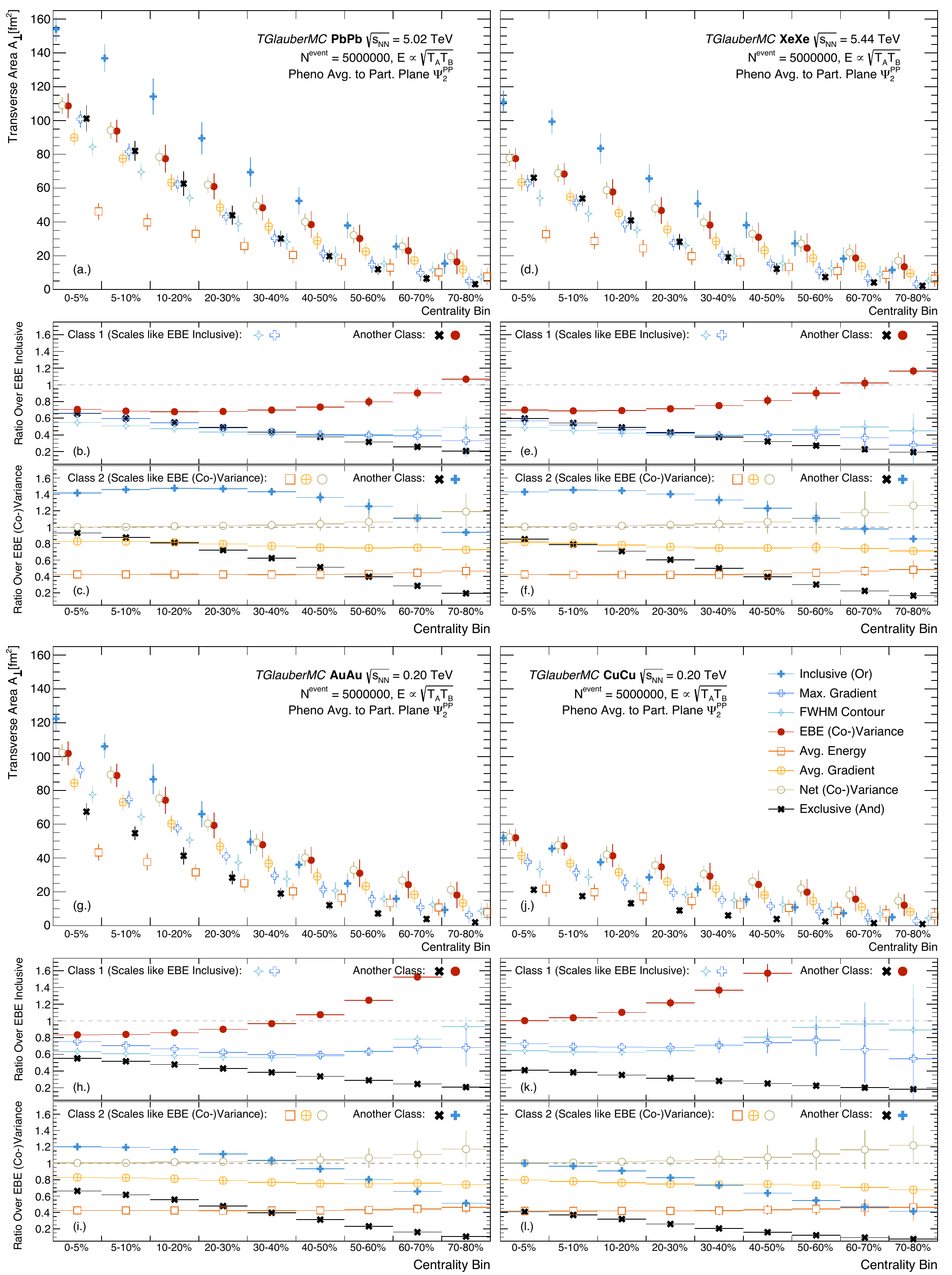} 
    \caption{Scalings of the various Glauber methods proposed to compute the transverse area across a range of collision species and energy. The panels are \PbPb\ at \sNN\ = 5.02 TeV (a.\textendash c.), \XeXe\ at \sNN\ = 5.44 TeV (d.\textendash f.), \AuAu\ at \sNN\ = 200 GeV (g.\textendash i.), and \CuCu\ at \sNN\ = 200 GeV (j.\textendash l.). Each column contains the explicit transverse areas \Aperp\ produced by each method (a., d., g., j.), ratios against the EBE inclusive \Ainc\ (b., e., h., k.) and ratios against the EBE width area \Awid\ (c., f., i., l.). EBE calculations produced by the Glauber code~\cite{Loizides:2017ack} are shown in solid markers while phenomenological area methods are shown in hollow markers.}
    \label{fig:appA:all_areas}
\end{figure*}

 Figure~\ref{fig:appA:all_areas} provides evidence that the categorization of area definitions into the \Awid\ and \Ainc\ classes holds across collision systems studied here. The figure shows the Glauber area calculations for \PbPb\ collisions at \sNN\ = 5.02 TeV, \XeXe\ at \sNN\ = 5.44 TeV, and \AuAu\ and \CuCu\ at \sNN\ = 200 GeV as a function of centrality. In addition, the ratios against the EBE inclusive (\Ainc) and EBE width-based (\Awid) are shown for each collision system. Solid markers indicate the area EBE calculations produced by the Glauber code~\cite{Loizides:2017ack} and hollow markers represent those from the phenomenological edge-area methods. For each system the ratio plots reveal the same two classes plus one outlier, the EBE exclusive (\Aexc). These observations coincide with the same classifications in \PbPb\ at \sNN\ = 2.76 TeV, presented in Fig.~\ref{fig:areascale_PbPb2.76}. Members of the same class are identified via their similar trends as a function of centrality and hence flat ratios. Throughout the plotting in Figs.~\ref{fig:areascale_PbPb2.76} and~\ref{fig:appA:all_areas}, members of the same class share similar colors and marker shapes to better delineate them visually. 

Table~\ref{tab:A:edensity_inputs_combined} details the experimental data used to estimate the mid-rapidity initial energy densities \ebj: the Jacobian to translate from \dnchdeta\ to \ensuremath{\rmd N_{\rm ch}/\rmd y}, and the transverse areas \Awid\ and \Ainc. Also included are the MSE/DoF fit metric results for the \Deltapt\ shifted spectra, as measured against the \TAA-scaled Tsallis \pp\ reference. Note that for ATLAS \XeXe~\cite{ATLAS:2022kqu} and PHENIX \AuAu~\cite{PHENIX:2003djd} the mid-rapidity \dnchdeta\ from ALICE~\cite{ALICE:2018hza} and STAR~\cite{STAR:2009mwg} respectively were used for \ebj\ computation, as published ATLAS/PHENIX \dnchdeta\ data in the same centrality binning were not available.

The averaged event harmonics, $c_0$, $c_2$ and the ratio $c_2/c_0$, from the Glauber simulated \Awid\ and \Ainc\ areas for the collision systems studied are presented in Table~\ref{tab:B:glauber_shape_harmonics_combined}. Note that the Average Radius method represents the \Awid\ class, while the Energy Half-Max Contour is in the \Ainc\ class. These values are used in the model estimate for high-\pT\ hadron elliptic flow \vtwo.

\begin{table*}[h!]
    \centering
    \renewcommand{\arraystretch}{1.3}
    \begin{tabular}{|c|c|c|c|c|c|c|c|}
        \hline
        \makecell{Centrality} 
        & \multirow{2}{1.5cm}{\centering $\displaystyle{\frac{\rmd N_{\rm ch}}{\rmd \eta}}$} 
        & \multicolumn{2}{c|}{Transverse Area (fm\textsuperscript{2})}
        & \multicolumn{2}{c|}{Energy Density \ebj\ (GeV/fm\textsuperscript{3})}  
        & \multicolumn{2}{c|}{\Deltapt\ Best-Fit Metric} \\ 
        \cline{3-6}
        Bin (\%) & & Width \Awid & Inclusive \Ainc & \hspace{6mm}Width\hspace{6mm} & Inclusive & \multicolumn{2}{c|}{MSE/DoF, Eq.~\eqref{eq:MSE_definition}}\\
        \hline \hline
        \multicolumn{6}{|l|}{\hspace{0.3cm}\PbPb\ at \sNN\ = 2.76 TeV. Jacobian $J = 1.09$ from~\cite{CMS:2012}, \dnchdeta\ from~\cite{ALICE:2010mlf}.} & \multicolumn{2}{c|}{ALICE~\cite{ALICE:2018vuu}} \\
        \hline
        0.0–5.0   & $1601 \pm 60$ & $107.56 \pm 7.04$ & $149.18 \pm 7.51$ & $71.80 \pm 4.60$ & $46.42 \pm 2.97$ & \multicolumn{2}{c|}{0.033} \\
        5.0–10.0  & $1294 \pm 49$ & $93.21 \pm 7.04$  & $131.96 \pm 8.26$ & $65.42 \pm 4.21$ & $41.16 \pm 2.65$ & \multicolumn{2}{c|}{0.028} \\
        10.0–20.0 & $966 \pm 37$  & $77.34 \pm 8.42$  & $109.77 \pm 10.61$ & $56.83 \pm 3.69$ & $35.63 \pm 2.31$ & \multicolumn{2}{c|}{0.021} \\
        20.0–30.0 & $649 \pm 23$  & $61.24 \pm 7.92$  & $85.57 \pm 9.46$  & $45.65 \pm 2.83$ & $29.22 \pm 1.81$ & \multicolumn{2}{c|}{0.013} \\
        30.0–40.0 & $426 \pm 15$  & $48.98 \pm 7.92$  & $65.91 \pm 8.65$  & $35.07 \pm 2.16$ & $23.61 \pm 1.46$ & \multicolumn{2}{c|}{0.013} \\
        40.0–50.0 & $261 \pm 9$   & $39.15 \pm 8.17$  & $49.34 \pm 7.98$  & $24.60 \pm 1.50$ & $18.08 \pm 1.10$ & \multicolumn{2}{c|}{0.014} \\
        50.0–60.0 & $149 \pm 6$   & $30.98 \pm 8.42$  & $35.27 \pm 7.40$  & $15.92 \pm 1.07$ & $13.39 \pm 0.90$ & \multicolumn{2}{c|}{0.023} \\
        60.0–70.0 & $76 \pm 4$    & $23.69 \pm 8.42$  & $23.34 \pm 6.82$  & $9.28 \pm 0.75$  & $9.47 \pm 0.76$  & \multicolumn{2}{c|}{0.009} \\
        70.0–80.0 & $32 \pm 2$    & $17.03 \pm 7.67$  & $13.91 \pm 5.83$  & $4.55 \pm 0.42$  & $5.95 \pm 0.55$  & \multicolumn{2}{c|}{0.025} \\
        \hline \hline
        \multicolumn{6}{|l|}{\hspace{0.3cm}\PbPb\ at \sNN\ = 5.02 TeV. Jacobian $J = 1.09$ from~\cite{CMS:2012}, \dnchdeta\ from~\cite{ALICE:2018vuu}.} & \multicolumn{2}{c|}{ALICE~\cite{ALICE:2018vuu}} \\
        \hline
        0.0–5.0   & $1910 \pm 49$ & $107.47 \pm 7.16$ & $153.41 \pm 7.42$  & $91.11 \pm 4.80$ & $56.68 \pm 2.98$ & \multicolumn{2}{c|}{0.058} \\
        5.0–10.0  & $1547 \pm 40$ & $93.04 \pm 7.04$  & $136.45 \pm 8.45$  & $83.37 \pm 4.40$ & $50.03 \pm 2.64$ & \multicolumn{2}{c|}{0.046} \\
        10.0–20.0 & $1180 \pm 31.6$ & $77.07 \pm 8.42$ & $114.01 \pm 10.87$ & $74.68 \pm 4.00$ & $44.30 \pm 2.38$ & \multicolumn{2}{c|}{0.032} \\
        20.0–30.0 & $786 \pm 20.8$  & $60.95 \pm 7.92$  & $89.40 \pm 9.80$  & $59.41 \pm 3.17$ & $35.64 \pm 1.90$ & \multicolumn{2}{c|}{0.021} \\
        30.0–40.0 & $512 \pm 15.5$  & $48.62 \pm 7.92$  & $69.22 \pm 8.97$  & $45.34 \pm 2.58$ & $28.31 \pm 1.61$ & \multicolumn{2}{c|}{0.014} \\
        40.0–50.0 & $318 \pm 12.5$  & $38.78 \pm 8.17$  & $52.20 \pm 8.37$  & $32.49 \pm 2.14$ & $21.85 \pm 1.44$ & \multicolumn{2}{c|}{0.006} \\
        50.0–60.0 & $183 \pm 8.2$   & $30.54 \pm 8.42$  & $37.57 \pm 7.79$  & $21.38 \pm 1.54$ & $16.22 \pm 1.17$ & \multicolumn{2}{c|}{0.004} \\
        60.0–70.0 & $96 \pm 5.9$    & $23.24 \pm 8.42$  & $25.12 \pm 7.21$  & $13.02 \pm 1.19$ & $11.74 \pm 1.07$ & \multicolumn{2}{c|}{0.014} \\
        70.0–80.0 & $45 \pm 3.5$    & $16.57 \pm 7.54$  & $15.12 \pm 6.25$  & $7.44 \pm 0.83$  & $8.41 \pm 0.93$  & \multicolumn{2}{c|}{0.010} \\
        \hline \hline
        \multicolumn{6}{|l|}{\hspace{0.3cm}\XeXe\ at \sNN\ = 5.44 TeV. Jacobian $J = 1.09$ from~\cite{CMS:2012}, \dnchdeta\ from~\cite{ALICE:2018hza}.} & ALICE~\cite{ALICE:2018hza} & ATLAS~\cite{ATLAS:2022kqu} \\
        \hline
        0.0-5.0   & $1167 \pm 26$ & $77.47 \pm 6.16$  & $110.87 \pm 6.77$  & $73.07 \pm 3.64$  & $45.31 \pm 2.26$ & 0.032 & 0.044 \\
        5.0-10.0  & $939 \pm 24$  & $68.33 \pm 6.41$  & $99.32 \pm 7.42$   & $64.65 \pm 3.40$  & $39.27 \pm 2.06$ & 0.054 & 0.059 \\
        10.0-20.0 & $706 \pm 17$  & $57.72 \pm 7.54$  & $83.49 \pm 9.05$   & $55.35 \pm 2.84$  & $33.84 \pm 1.74$ & 0.037 & 0.027 \\
        20.0-30.0 & $478 \pm 11$  & $46.78 \pm 7.79$  & $65.66 \pm 8.52$   & $43.56 \pm 2.20$  & $27.71 \pm 1.40$ & 0.030 & 0.013 \\
        30.0-40.0 & $315 \pm 8$   & $38.16 \pm 8.17$  & $50.81 \pm 8.14$   & $32.77 \pm 1.72$  & $22.37 \pm 1.17$ & 0.018 & 0.007 \\
        40.0-50.0 & $198 \pm 5$   & $30.96 \pm 8.55$  & $38.11 \pm 7.80$   & $23.32 \pm 1.22$  & $17.68 \pm 0.92$ & 0.040 & 0.007 \\
        50.0-60.0 & $118 \pm 3$   & $24.55 \pm 8.80$  & $27.26 \pm 7.47$   & $15.93 \pm 0.84$  & $13.86 \pm 0.73$ & 0.019 & 0.005 \\
        60.0-70.0 & $64.7 \pm 2$  & $18.65 \pm 8.29$  & $18.28 \pm 6.86$   & $10.32 \pm 0.59$  & $10.60 \pm 0.61$ & 0.019 & - \\
        70.0-80.0 & $32 \pm 1.3$  & $13.48 \pm 7.16$  & $11.57 \pm 5.64$   & $6.22 \pm 0.42$   & $7.62 \pm 0.51$  & 0.091 & - \\
        60.0-80.0 & $48 \pm  1.65$ & $16.07 \pm 8.55$ & $14.92 \pm 7.12$ & $8.47 \pm 0.51 $ & $9.35 \pm 0.56$ & - & 0.004 \\
        \hline \hline
        \multicolumn{6}{|l|}{\hspace{0.3cm}\AuAu\ at \sNN\ = 200 GeV. Jacobian $J = 1.25$ from~\cite{PHENIX:2015tbb}, \dnchdeta\ from~\cite{BRAHMS:2001llo}.} & STAR~\cite{STAR:2003fka,STAR:2011iap} & PHENIX~\cite{PHENIX:2003djd} \\
        \hline
        0.0–5.0   & $625 \pm 55$  & $101.84 \pm 7.04$ & $122.34 \pm 7.31$ & $24.37 \pm 3.02$ & $19.08 \pm 2.36$ & 0.109 & 0.237 \\
        5.0–10.0  & $501 \pm 44$  & $88.78  \pm 6.79$ & $106.01 \pm 7.13$ & $21.77 \pm 2.69$ & $17.19 \pm 2.13$ & 0.171 & - \\
        0.0–10.0  & $563 \pm 55$ & $95.31 \pm 7.67$ & $114.17 \pm 8.32$ & $23.14 \pm 2.58$ & $ 18.18 \pm 2.03$ & - & 0.129 \\
        10.0–20.0 & $377 \pm 33$  & $74.16  \pm 8.04$ & $86.56  \pm 8.91$ & $18.94 \pm 2.34$ & $15.41 \pm 1.90$ & 0.098 & 0.058 \\
        20.0–30.0 & $257 \pm 23$  & $59.28  \pm 7.54$ & $65.97  \pm 7.79$ & $15.32 \pm 1.93$ & $13.28 \pm 1.67$ & 0.086 & 0.047 \\
        30.0–40.0 & $174 \pm 16$  & $47.83  \pm 7.67$ & $49.51  \pm 7.08$ & $12.12 \pm 1.56$ & $11.58 \pm 1.49$ & 0.140 & 0.008\\
        \hline \hline
        \multicolumn{6}{|l|}{\hspace{0.3cm}\CuCu\ at \sNN\ = 200 GeV. Jacobian $J = 1.25$ from~\cite{PHENIX:2015tbb}, \dnchdeta\ from~\cite{STAR:2009mwg}.} & \multicolumn{2}{c|}{STAR~\cite{STAR:2009mwg}} \\
        \hline
        0.0–10.0 & $176.3 \pm 12.7$ & $49.57 \pm 6.16$ & $48.69 \pm 5.20$  & $11.76 \pm 1.22$ & $12.05 \pm 1.25$ & \multicolumn{2}{c|}{0.010}  \\
        10.0–20.0& $121.5 \pm 8.7$  & $41.29 \pm 6.66$ & $37.49 \pm 4.83$  & $9.14 \pm 0.95$  & $10.39 \pm 1.08$ & \multicolumn{2}{c|}{0.006} \\
        20.0–40.0& $69.6 \pm 4.9$   & $31.95 \pm 8.04$ & $24.96 \pm 5.69$  & $6.12 \pm 0.62$  & $8.50 \pm 0.87$  & \multicolumn{2}{c|}{0.0004} \\
        \hline
    \end{tabular}
    \caption{Inputs for the energy density (mid-rapidity yields \dnchdeta, Jacobian $J$ and transverse area \Aperp\ for width and inclusive classes), resultant values of the energy density \ebj, and MSE/DoF fit metric results for the \Deltapt\ spectra shifts. Note that ALICE and STAR \dnchdeta\ results were used for ATLAS and PHENIX \ebj\ computation respectively. The errors quoted for the area calculations correspond to the standard deviation of the area within a given bin. However, the standard error on the mean is instead used for propagation to our final uncertainty on the energy density.}
    \label{tab:A:edensity_inputs_combined}
\end{table*}

\clearpage

\begin{table*}[h!]
    \centering
    \renewcommand{\arraystretch}{1.4}
    \begin{tabular}{|p{3cm}|p{2.7cm}|c|ccccccccc|}
        \hline
        \multirow{2}{3cm}{Collision System} & \multirow{2}{2cm}{Area Method and Class} & \multirow{2}{1cm}{Edge Coef.}
        & \multicolumn{9}{c|}{Centrality Bin (\%)} \\
        & & & 0–5\% & 5-10\% & 10-20\% & 20-30\% & 30-40\% & 40-50\% & 50-60\% & 60-70\% & 70-80\% \\
        \hline \hline
        \multirow{6}{3cm}{\PbPb\ 2.76 TeV} & \multirow{3}{2.5cm}{Avg. Radius (\Awid)} & $c_0$ & 3.83 & 3.55 & 3.23 & 2.85 & 2.53 & 2.25 & 2.00 & 1.77 & 1.53  \\
        & & $c_2$ & 0.14 & 0.26 & 0.38 & 0.48 & 0.53 & 0.55 & 0.56 & 0.54 & 0.50  \\
        \cline{3-12}
        & & $c_2/c_0$ & 0.03 & 0.07 & 0.12 & 0.17 & 0.21 & 0.24 & 0.28 & 0.31 & 0.32  \\
        \cline{2-12}
        & \multirow{3}{2.7cm}{FWHM Contour (\Ainc)} & $c_0$ & 5.17 & 4.70 & 4.13 & 3.49 & 2.96 & 2.51 & 2.15 & 1.87 & 1.49  \\
        & & $c_2$ & 0.32 & 0.47 & 0.62 & 0.73 & 0.77 & 0.75 & 0.73 & 0.72 & 0.68  \\
        \cline{3-12}
        & & $c_2/c_0$ & 0.06 & 0.10 & 0.15 & 0.21 & 0.26 & 0.30 & 0.34 & 0.38 & 0.45  \\
        \hline \hline
        \multirow{6}{3cm}{\PbPb\ 5.02 TeV} & \multirow{3}{2.5cm}{Avg. Radius (\Awid)} & $c_0$ & 3.83 & 3.55 & 3.22 & 2.84 & 2.52 & 2.24 & 1.99 & 1.75 & 1.51  \\
        & & $c_2$ & 0.14 & 0.27 & 0.38 & 0.48 & 0.53 & 0.56 & 0.56 & 0.54 & 0.50  \\
        \cline{3-12}
        & & $c_2/c_0$ & 0.03 & 0.07 & 0.12 & 0.17 & 0.21 & 0.25 & 0.28 & 0.31 & 0.33  \\
        \cline{2-12}
        & \multirow{3}{2.7cm}{FWHM Contour (\Ainc)} & $c_0$ & 5.17 & 4.69 & 4.12 & 3.48 & 2.94 & 2.50 & 2.14 & 1.84 & 1.45  \\
        & & $c_2$ & 0.31 & 0.47 & 0.62 & 0.73 & 0.77 & 0.75 & 0.72 & 0.71 & 0.67  \\
        \cline{3-12}
        & & $c_2/c_0$ & 0.06 & 0.10 & 0.15 & 0.21 & 0.26 & 0.30 & 0.34 & 0.38 & 0.46  \\
        \hline \hline
        \multirow{6}{3cm}{\XeXe\ 5.44 TeV} & \multirow{3}{2.5cm}{Avg. Radius (\Awid)} & $c_0$ & 3.22 & 3.02 & 2.77 & 2.48 & 2.24 & 2.02 & 1.82 & 1.62 & 1.39  \\
        & & $c_2$ & 0.18 & 0.24 & 0.31 & 0.39 & 0.44 & 0.47 & 0.47 & 0.46 & 0.44  \\
        \cline{3-12}
        & & $c_2/c_0$ & 0.05 & 0.08 & 0.11 & 0.15 & 0.19 & 0.23 & 0.26 & 0.28 & 0.31  \\
        \cline{2-12}
        & \multirow{3}{2.7cm}{FWHM Contour (\Ainc)} & $c_0$ & 4.13 & 3.77 & 3.32 & 2.85 & 2.47 & 2.17 & 1.94 & 1.62 & 1.20  \\
        & & $c_2$ & 0.34 & 0.40 & 0.50 & 0.57 & 0.59 & 0.61 & 0.65 & 0.67 & 0.59  \\
        \cline{3-12}
        & & $c_2/c_0$ & 0.08 & 0.10 & 0.15 & 0.20 & 0.24 & 0.28 & 0.33 & 0.41 & 0.49  \\
        \hline \hline
        \multirow{6}{3cm}{\AuAu\ 0.20 TeV} & \multirow{3}{2.5cm}{Avg. Radius (\Awid)} & $c_0$ & 3.70 & 3.45 & 3.15 & 2.80 & 2.51 & 2.25 & 2.02 & 1.80 & 1.58  \\
        & & $c_2$ & 0.16 & 0.26 & 0.36 & 0.45 & 0.50 & 0.53 & 0.54 & 0.53 & 0.50  \\
        \cline{3-12}
        & & $c_2/c_0$ & 0.04 & 0.07 & 0.11 & 0.16 & 0.20 & 0.23 & 0.26 & 0.29 & 0.31  \\
        \cline{2-12}
        & \multirow{3}{2.7cm}{FWHM Contour (\Ainc)} & $c_0$ & 4.95 & 4.51 & 3.99 & 3.40 & 2.91 & 2.51 & 2.18 & 1.92 & 1.57  \\
        & & $c_2$ & 0.33 & 0.45 & 0.58 & 0.69 & 0.73 & 0.72 & 0.70 & 0.70 & 0.67  \\
        \cline{3-12}
        & & $c_2/c_0$ & 0.06 & 0.10 & 0.14 & 0.20 & 0.25 & 0.28 & 0.32 & 0.36 & 0.42  \\
        \hline \hline
        \multirow{6}{3cm}{\CuCu\ 0.20 TeV} & \multirow{3}{2.5cm}{Avg. Radius (\Awid)} & $c_0$ & 2.63 & 2.50 & 2.33 & 2.13 & 1.96 & 1.80 & 1.64 & 1.47 & 1.29  \\
        & & $c_2$ & 0.18 & 0.22 & 0.27 & 0.33 & 0.38 & 0.40 & 0.40 & 0.39 & 0.39  \\
        \cline{3-12}
        & & $c_2/c_0$ & 0.07 & 0.09 & 0.11 & 0.15 & 0.19 & 0.22 & 0.24 & 0.26 & 0.30  \\
        \cline{2-12}
        & \multirow{3}{2.7cm}{FWHM Contour (\Ainc)} & $c_0$ & 3.24 & 3.00 & 2.71 & 2.39 & 2.14 & 1.94 & 1.71 & 1.41 & 1.11  \\
        & & $c_2$ & 0.35 & 0.39 & 0.42 & 0.47 & 0.52 & 0.58 & 0.63 & 0.64 & 0.54  \\
        \cline{3-12}
        & & $c_2/c_0$ & 0.10 & 0.13 & 0.15 & 0.19 & 0.24 & 0.30 & 0.36 & 0.45 & 0.49  \\
        \hline

    \end{tabular}
    \caption{Tabulated values for averaged event harmonics $c_2,\, c_0$ from the \Awid\ and \Ainc\ areas generated from Glauber simulations for the collision systems studied. Note that the Average Radius method represents the \Awid\ class, while the Energy Half-Max Contour represents in the \Ainc\ class.}
    \label{tab:B:glauber_shape_harmonics_combined}
\end{table*}


\newpage
\section{Exploration of Fitting Procedures and Fit Metrics for \pT\ Spectrum Fitting}
\label{appendix:fitting_methods}
At first glance, the choice of the MSE fit metric Eq.~\eqref{eq:MSE_definition} over conventional choices like the $\chi^2$ or Likelihood appears arbitrary. This appendix provides details on the studies undertaken to select the MSE metric, where it performs better than the $\chi^2$ and Likelihood in extracting parameters of interest from \pT\ spectra, particularly the high \pT\ power law tail. 

To begin, it is helpful to compare the MSE more directly to the $\chi^2$ from their respective definitions. The $\chi^2$ between an observed dataset $O_i$ and fitting expectation $E_i$ with corresponding error $\delta E_i$ is given by
\begin{equation}
    \chi^2 = \sum_{\mathrm{bins} \ i} \left(\frac{O_i - E_i}{\delta E_i}\right)^2.
\end{equation}
The MSE metric, defined by Eq.~\eqref{eq:MSE_definition}, admits the power series expansion in the case of reasonable fits $|O_i / E_i| < 2$
\begin{align}
    \mathrm{MSE} &= \sum_{\mathrm{bins} \ i} \left(\frac{O_i - E_i}{E_i} - \frac{1}{2} \left(\frac{O_i - E_i}{E_i}\right)^2 + \cdots \right)^2 \nonumber \\
    &= \sum_{\mathrm{bins} \ i} \left(\frac{O_i - E_i}{E_i} \right)^2 + \mathcal{O}\Big( (O_i - E_i) / E_i \Big)^4. \label{eq:MSE_chi2_limit}
\end{align}
That is to say, an MSE fit is at leading order in $(O_i - E_i) / E_i$ exactly a $\chi^2$ fit with error $\delta E_i \propto E_i$, a generous but reasonable assumption for measured \pT\ spectra for which statistical and experimental systematic uncertainties are folded together. From this point of view, it's clear that the MSE metric is bounded below by the corresponding $\chi^2$ for the same fit, which assures that the fits discussed in Tabs.~\ref{table:ppMSE}, ~\ref{tab:A:edensity_inputs_combined} are of quality as good or better than for $\chi^2 / \mathrm{DoF}$ equivalent to the resulting MSE.

\begin{figure}
    \centering
    \includegraphics[width=0.99\linewidth]{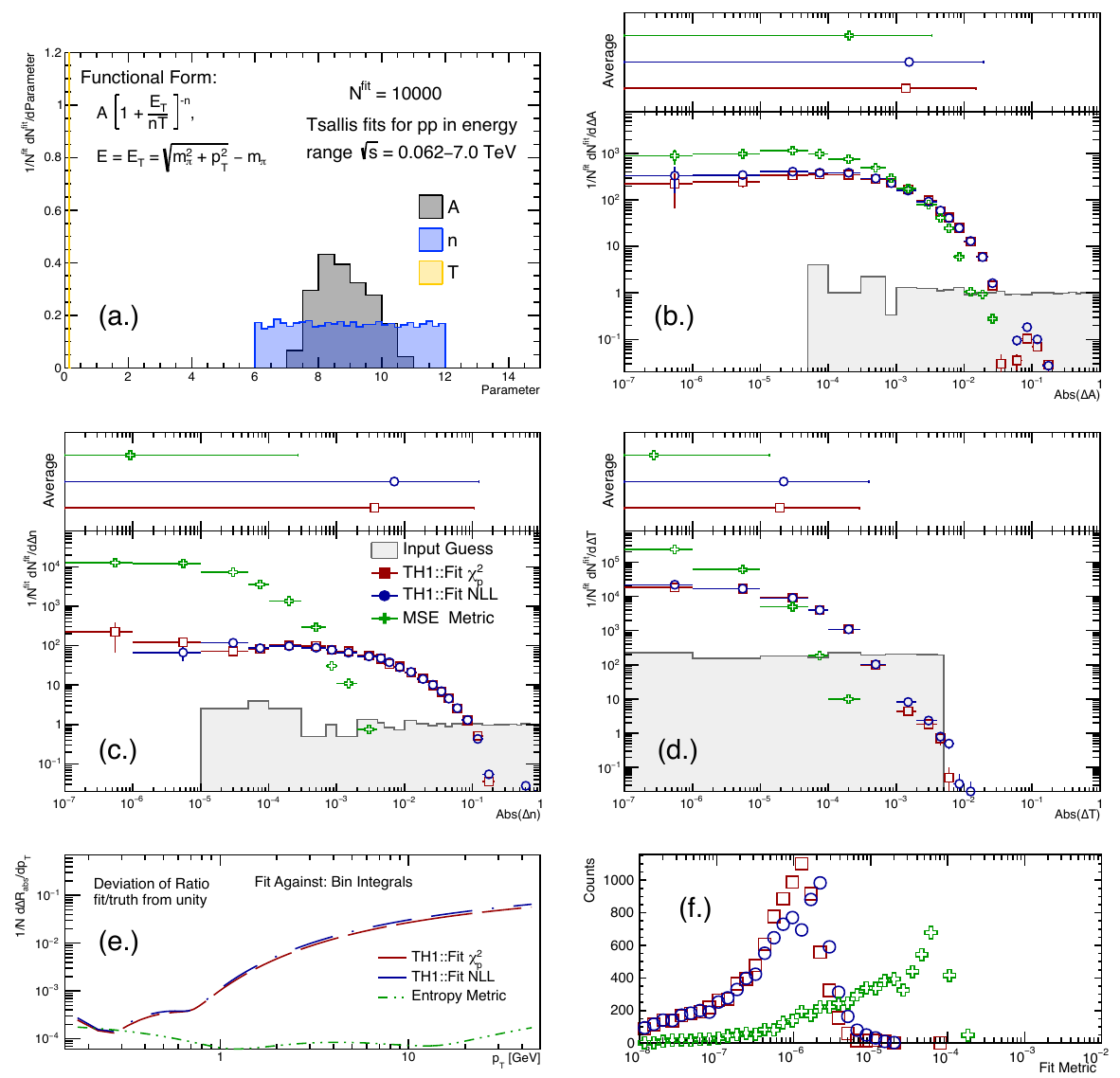}
    \caption{Comparison of fitting results for the Pearson $\chi^2$, Likelihood and MSE metrics for toy Tsallis fits against exact Tsallis distributions. Panel (a.) shows the distribution of Tsallis functions used to generate toy data over the 10,000 fits. Panels (b.), (c.) and (d.) show the distribution of resultant fit parameters $A,n,T$ respectively, relative to the thrown true parameters, as well as the uniformly scrambled inputs given to the fitter. Panel (e.) shows the average deviation of the ratio fit/truth from unity over the \pT\ range for the different fitting methods, and panel (f.) displays the distributions of minimized fit metric values for each method.}
    \label{fig:appendix_fitting_comparison}
\end{figure}

This error scaling additionally produces a unitless fit metric, and weights all bins $i$ equally against the fit form. This reduces the bias toward bins with larger magnitude, which bias traditional $\chi^2$ or Likelihood fits against the correct parameters for steeply falling power law spectra. To explore this quantitatively, we consider Tsallis distributions of the form Eq.~\eqref{eq:Tsallis}, with parameters in ranges $n \in [6,12],\ T \in [0.1, 0.25]$ GeV, and normalization $A$ chosen so that the distribution integrates to unity over the range $\pT \in [5, 50]$ GeV. The ranges are chosen using fit parameters from Refs.~\cite{Wong:2012zr,Wilk:1999dr,ATLAS:2010jvh,ALICE:2013txf,STAR:2006nmo,PHENIX:2011rvu}, and roughly capture the range of fits obtained against \pp\ data spectra in this study, shown in Tab.~\ref{table:ppMSE}. The distributions of input Tsallis parameters can be seen in panel (a.) of Fig.~\ref{fig:appendix_fitting_comparison}.

These Tsallis distributions are then sampled to fill histograms in the same binning as in \pp\ data spectra above the relevant \pT\ threshold used in our analysis, either to the same statistics as in the data or to higher statistics. Then, the parameters are shuffled in a wide range, and passed in as initial guesses to ROOT's TMinuit Migrad fitter~\cite{Fletcher:1970fit, James:1975dr}, set to use $\chi_{\rm p}^2$, Negative Log Likelihood or MSE fitting as the goodness-of-fit metric. The resultant parameters from fits are then compared against the thrown truth parameters, and the difference $|\{A,n,T\}_{\rm thrown} - \{A,n,T\}_{\rm fitted}|$ is tabulated in a histogram for each fit, which can be seen in panels (b.-d.) of Fig.~\ref{fig:appendix_fitting_comparison}. In addition, the average absolute residual of the fit against the thrown distribution is tabulated in Fig.~\ref{fig:appendix_fitting_comparison} (e.) across the \pT\ range to study the bias of fitting methods against \pT. Finally, the distribution of resultant minimized fit metrics $\chi_{\rm p}^2,\  \chi_{\rm NLL}^2, $ and MSE are shown for the fits in Fig.~\ref{fig:appendix_fitting_comparison} (f.) to compare the relative magnitudes for successful fits. A total of 10,000 sets of Tsallis parameters were generated and fit for this study.

Note that all fits are performed by comparing the sampled bin contents against the integral of the fit form in that bin. In the specific case shown in Fig.~\ref{fig:appendix_fitting_comparison}, the histograms are extrapolated to infinite statistics, i.e. filled directly with the integral of the truth distribution, so that in principle exact parameters could be found by the fitter. This case most strongly exhibits the biases present in $\chi^2$ and Likelihood fits; despite full access to exact truth information, these methods still contain a strong bias against the correct parameters due to over-weighting bins at low \pT. However, it is worth noting that similar behaviors were observed in fits with finite samplings, and in the case of samplings according to the statistics observed in corresponding datasets. The infinite statistics case is shown here for simplicity and brevity, as each \pp\ distribution studied here and fit in Tab.~\ref{table:ppMSE} contains varying counts of hadrons.

The bias toward low \pT\ in the $\chi^2$ and NLL fitters is most apparent in Fig.~\ref{fig:appendix_fitting_comparison} (e.), where the residuals in the ratio fit/truth explode at high \pT. Meanwhile, the MSE fit remains flat over the full range, indicating that it weights the full \pT\ range equally. As a consequence, the resultant MSE fits extract the power law $n$ with two orders of magnitude smaller error on average than the $\chi^2$ and NLL, though the improvements for the remaining parameters $A, \ T$ are smaller and likely a direct consequence of better extraction of $n$. This precision extraction of the power law motivates the MSE for this study, where a good extraction of the power law tail is essential to correctly estimate \Deltapt. 

The extracted MSE/DoF can be interpreted similarly to a $\chi^2$/DoF, as shown in Eq.~\eqref{eq:MSE_chi2_limit} and in Fig.~\ref{fig:appendix_fitting_comparison} (f.). Moreover, we found that when attempting to fit an incorrect form, the MSE explodes as expected for the $\chi^2$. To show this, we fit an exponential $A \pT e^{-B \pT + C}$ against the \pp\ spectra in the range $\pT \in [5, 50]$ GeV, which is of comparable order of magnitude in the fitting region but has an incorrect shape relative to the \pp\ spectra. Then, we fit the Tsallis form against data sampled from this fit exponential using the $\chi^2, $ NLL and MSE fits. We found that the resultant fit metrics per degree of freedom were all larger than 5, and within $\pm 3$ of one another, which demonstrates that the MSE/DoF is comparable to a $\chi^2$/DoF for bad fits as well as good ones.

\newpage
\section{Investigation of \Deltapt\ Fit Residuals for Subdominant Effects}
\label{appendix:deltapt_residuals}

The strong linear correlation observed in Fig.~\ref{fig:width-fit} merits further exploration. Specifically, it is worthwhile to check if the residuals to the linear fits in $\Deltapt(\ebj)$ obtained in Sec.~\ref{sec:Results} exhibit any clear dependence against other parameters such as $\Npart, \Aperp$ that might reveal a residual geometry or path-length component. Specifically, we consider an asymptotic expansion
\begin{equation}\label{eq:energyloss_expansion}
    \Deltapt = \alpha \ebj + F(L_{\rm eff}, \varepsilon_2, \Npart, \cdots),
\end{equation}
\begin{figure}[h]
    \centering
    \includegraphics[width=0.6\linewidth]{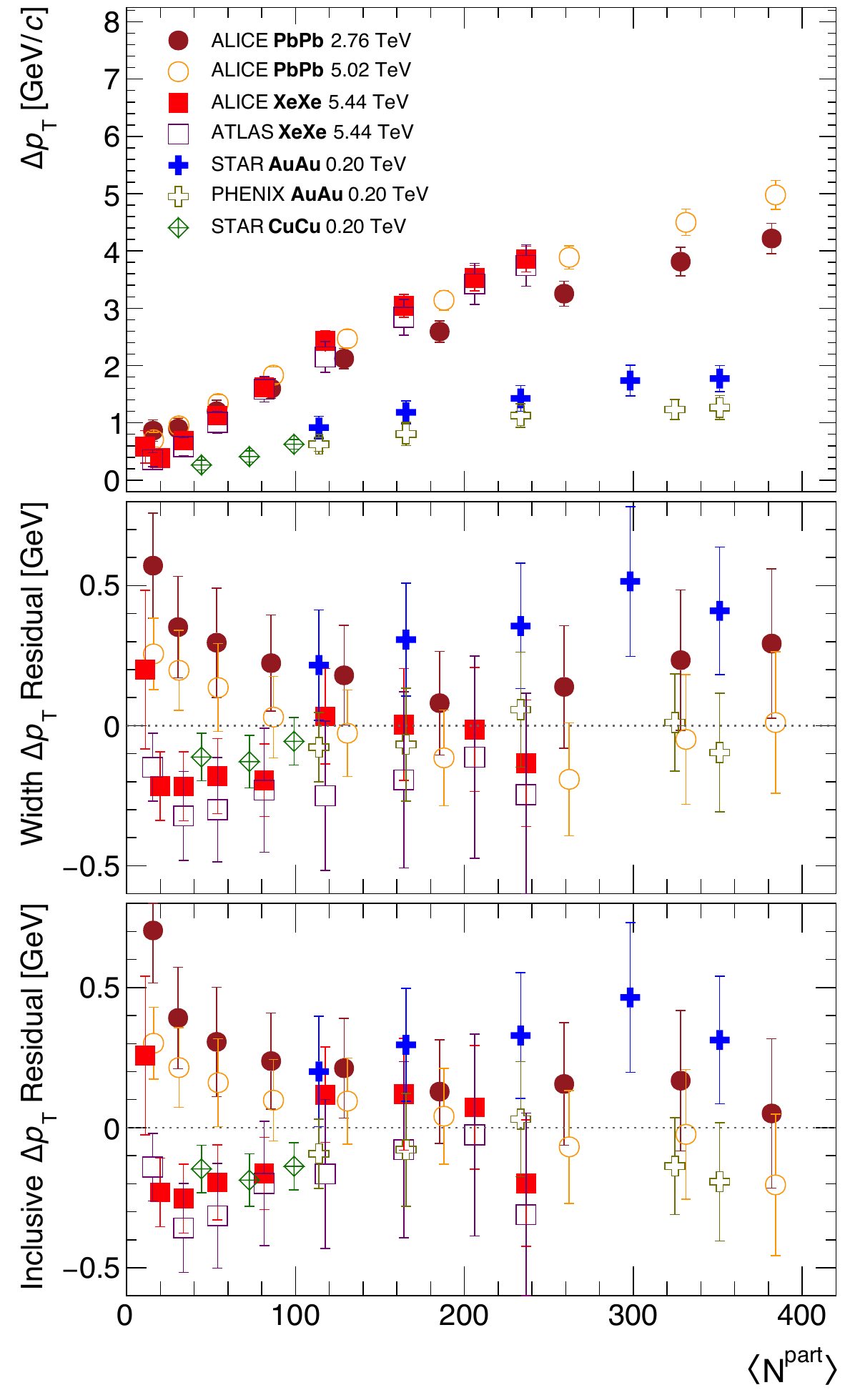}
    \caption{(Top) Energy loss $\Deltapt$ across \Npart\ for the various collision systems studied in this work. (Bottom) Residuals from the $\Deltapt(\ebj)$ linear fit plotted against \Npart. }
    \label{fig:eventshape_residuals_npart}
\end{figure}
with $\alpha = \mathrm{const.}$ and $F(L_{\rm eff}, \varepsilon_2, \Npart, \cdots) = o(\ebj)$ (little-O), and investigate the form of this additional contribution via the fit residuals
\begin{equation}\label{eq:energyloss_residuals}
    \delta(\Deltapt) = \Deltapt - \alpha \ebj = F(L_{\rm eff}, \varepsilon_2, \Npart, \cdots).
\end{equation}
Here $L_{\rm eff}$ denotes an effective medium path length and $\varepsilon_2$ an effective second order eccentricity/path-length difference scale $\Delta L_{\rm eff}$; these are expected to represent the strongest next-to-leading-order dependence though in principle $F$ may depend dynamically on a number of parameters.

The energy loss estimate $\Deltapt$ and corresponding residuals $\delta(\Deltapt)$ plotted against \Npart\ are shown in Fig.~\ref{fig:eventshape_residuals_npart}. Note that while \Deltapt\ and \Npart\ do not depend on the choice of area scaling method, the residuals $\delta(\Deltapt)$ do inherit a slight dependence on \Aperp\ from the fit form against \ebj, which varies slightly between \Aperp\ models as shown in Fig.~\ref{fig:width-fit}. Therefore, the residuals against both fits are shown in the breakout panels of Fig.~\ref{fig:width-fit}. The two are very similar, affirming that both the width based and inclusive models produce consistent linear fits, albeit with different slope.

In the \Npart\ dependence of $\Deltapt$ (top panel of Fig.~\ref{fig:eventshape_residuals_npart}), a clear separation between the LHC and RHIC datasets is present, and attributable to the expected $\sNN$ dependence of energy loss. This separation disappears in the residuals against the \ebj\ fit, indicating that the \sNN\ dependent evolution of \Deltapt\ can be well captured by \ebj. In particular, the ALICE \PbPb\ datasets at $\sNN = 2.76,\ 5.02$ TeV~\cite{ALICE:2018vuu} now sit between the \AuAu~\cite{STAR:2003fka,STAR:2011iap} and \CuCu~\cite{STAR:2009mwg} datasets at $\sNN = 200$ GeV recorded by STAR. Note that the  discrepancy between the STAR and PHENIX~\cite{PHENIX:2003djd} \AuAu\ datasets is due to differences in the \pp\ reference spectrum as discussed in Sec.~\ref{sec:Results}. In general, the residual data do not show a clear trend against \Npart, indicating that the system size dependence of \Deltapt\ is also generally well described by the \ebj\ fitting. 

\begin{figure}
    \centering
    \includegraphics[width=0.99\linewidth]{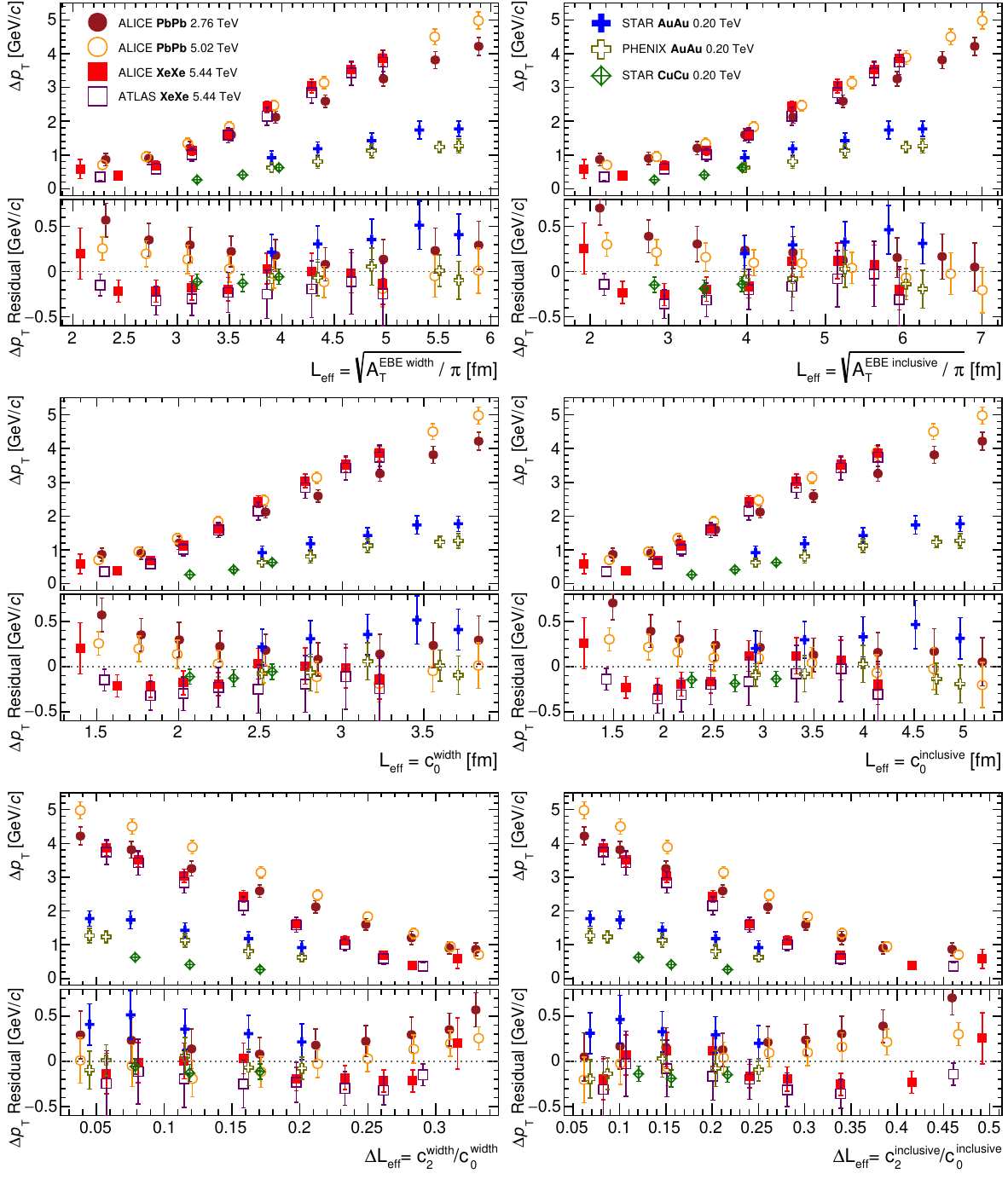}
    \caption{Energy loss $\Deltapt$ as a function of $L_{\rm eff}$ as modeled by $\sqrt{\Aperp / \pi}$ (first row) and the harmonic coefficient $c_0$ (second column), as well as the path length difference $\Delta L_{\rm eff} \sim c_2 / c_0$ for the various systems studied. For each such dependence, the residuals $\delta (\Deltapt)$ from the $\ebj$ fit are also shown in the corresponding breakout panel below. All plots in the left column correspond to the width-based area calculation \Awid, while those in the right use the inclusive area \Ainc.}
    \label{fig:eventshape_residuals}
\end{figure}

However, it can be seen in Fig.~\ref{fig:eventshape_residuals_npart} that the the larger systems (\AuAu, \PbPb) tend to sit above the smaller systems (\XeXe, \CuCu), which could indicate some dependence on nuclear shape, size or structure. To investigate these possibilities, the \Deltapt\ residuals are plotted against estimates of path length $L_{\rm eff}$ and path length difference $\Delta L_{\rm eff}$ in Fig.~\ref{fig:eventshape_residuals}. Both the EBE area and phenomenological methods are used to estimate $L_{\rm eff}$, using $L_{\rm eff} = \sqrt{\Aperp / \pi}$ for the EBE methods and $L_{\rm eff} = c_0$ for the phenomenological methods; for $\Delta L_{\rm eff}$, the phenomenological $c_2 / c_0$ is used. Results are shown for both the width class \Awid\ and the inclusive class \Ainc\ in the left and right columns of Fig.~\ref{fig:eventshape_residuals} respectively. Note that the coefficients $c_0, c_2$ come from the Average Radius method for the \Awid\ class, and from the Energy Half-Max Contour for the \Ainc\ class as is used above, and can be seen in Tab.~\ref{tab:B:glauber_shape_harmonics_combined}. The data generally show identical trends between both EBE and phenomenological estimates of the path length $L_{\rm eff}$, affirming the grouping of EBE and phenomenological methods into common classes performed in this study. In addition, the dependences on $\Delta L_{\rm eff}$ are reversed relative to those for $L_{\rm eff}$, consistent with a picture of collision geometry where the most central collisions possess the most isotropic overlap area, and thus the smallest path length difference.

Because all the residual plots share the same data in the $y$-axis, the overall ordering of datasets in the residuals is consistent among all plots of Fig.~\ref{fig:eventshape_residuals}. However, as for \Npart, the fitting against \ebj\ succeeds in capturing the general behavior of the datasets across both $L_{\rm eff}$ and $\Delta L_{\rm eff}$, as well as $\sNN$, leaving residuals in each case that do not exhibit a clear dependence on $\sqrt{\Aperp},\ c_0, $ or $c_2 / c_0$. Specifically, there is no clear leftover shape in the dependence of the residuals on $L_{\rm eff}$  or $\Delta L_{\rm eff}$ that follow a common trend among all systems studied. This strongly supports the claim made in the conclusions of Sec.~\ref{sec:Conc}: that the dominant behavior of \Deltapt\ can be well understood in terms of energy density alone, and further details of nuclear size, shape and collision geometry leave subdominant contributions to the partonic energy loss.

\newpage
\section{Additional $v_2$ predictions using $A_\cup$}
\label{Appendix:v2_inclusive}

\begin{figure}[h]
    \centering
    \includegraphics[width=0.5\linewidth]{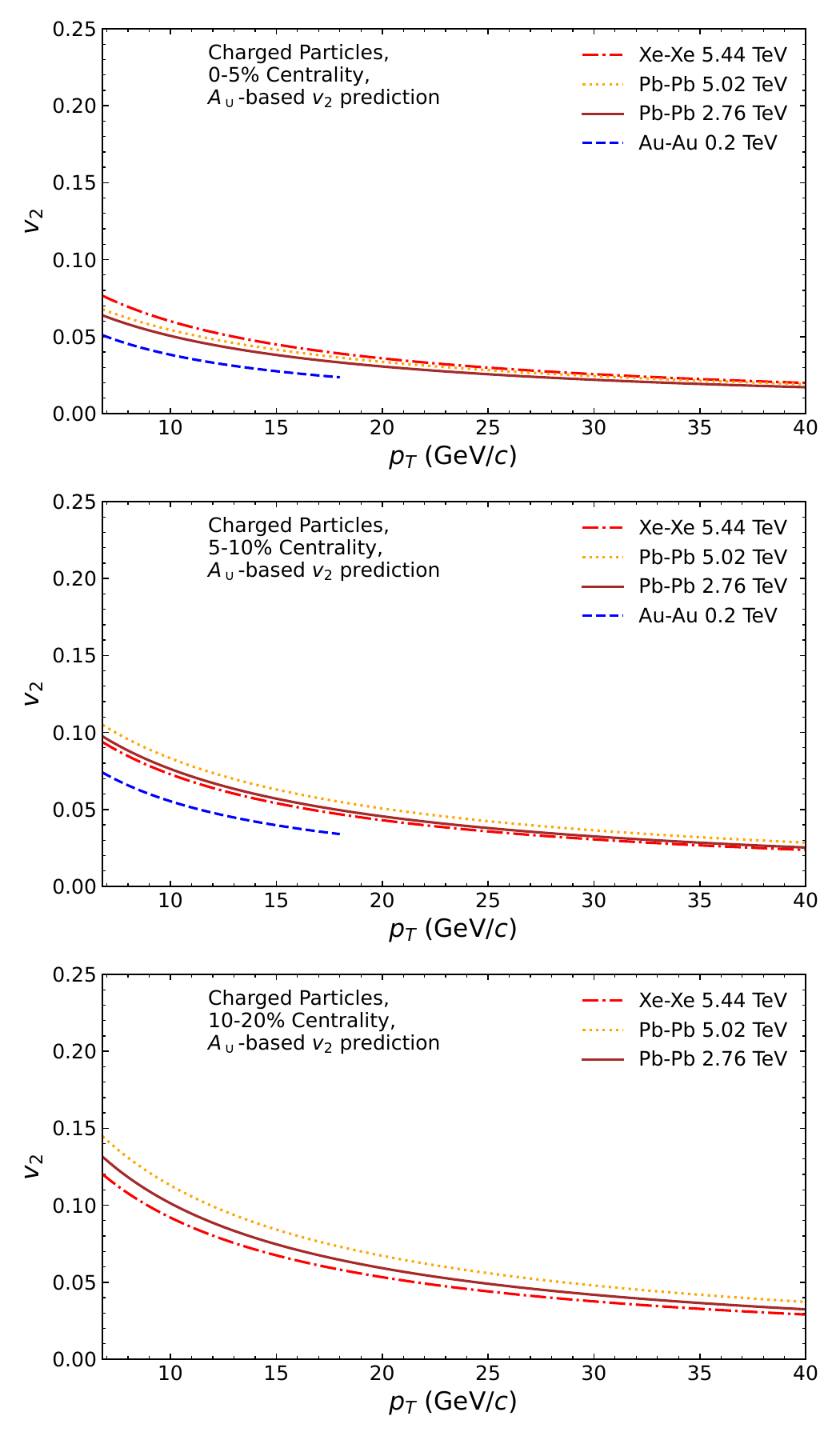}
    \caption{High \pT\ $v_2$ predictions using the \Deltapt\ and $A_\cup$ based model and  for a variety of collision systems.}
    \label{fig:v2_othersystems_inclusive_area_appendix}
\end{figure}

For the sake of completeness, and for comparison against the results using \Awid-based \vtwo\ shown in Fig.~\ref{fig:v2-OtherCollisions}, we also show the \vtwo\ using the \Ainc\ in Fig.~\ref{fig:v2_othersystems_inclusive_area_appendix}. 

\end{document}